%% file: main.tex
\documentclass[10pt,twocolumn,letterpaper]{article}

\usepackage[pagenumbers]{cvpr} 
\usepackage{comment}
\usepackage{lipsum}  
\newcommand*{\V}[1]{\boldsymbol{#1}}   
\newcommand*{\M}[1]{\mathbf{#1}}       
\newcommand\norm[1]{\left\lVert#1\right\rVert}
\usepackage{amsmath}
\usepackage{graphics}
\DeclareMathOperator*{\argmax}{arg\,max}
\DeclareMathOperator*{\argmin}{arg\,min}
\usepackage{stmaryrd}
\usepackage{booktabs}

\usepackage{gensymb}
\usepackage{todonotes}

\usepackage{mathtools}

\usepackage{multirow}
\usepackage{rotating}
\usepackage{algorithm}
\usepackage{algpseudocode}
\usepackage{graphicx}
\usepackage{amssymb}
\usepackage{etoolbox}
\makeatletter
\robustify\@latex@warning@no@line
\makeatother
\usepackage{authblk}
\usepackage{graphicx}

\DeclareMathOperator{\tr}{tr}


\definecolor{cvprblue}{rgb}{0.21,0.49,0.74}
\usepackage[pagebackref,breaklinks,colorlinks,allcolors=cvprblue]{hyperref}

\newcommand{\greencheck}{{\color{Green}\checkmark}}
\newcommand{\redcross}{{\color{red}\text{\sffamily X}}}

\def\paperID{16590} 
\def\confName{CVPR}
\def\confYear{2025}

\title{A New Statistical Model of Star Speckles for Learning to Detect and Characterize Exoplanets in Direct Imaging Observations }

\author[1]{\vspace{-2ex}Théo Bodrito \thanks{Corresponding author: \texttt{theo.bodrito@inria.fr}}}
\author[2]{Olivier Flasseur}
\author[3]{Julien Mairal}
\author[1, 4]{Jean Ponce}
\author[2]{\\ Maud Langlois}
\author[5, 6]{Anne-Marie~Lagrange}

{
\affil[1]{\small Département d’Informatique de l’École normale supérieure (ENS-PSL, CNRS, Inria)}
\affil[2]{Universite Claude Bernard Lyon 1, Centre de Recherche Astrophysique de Lyon UMR 5574, \newline{}ENS de Lyon, CNRS, Villeurbanne, F-69622, France}
\affil[3]{Université Grenoble Alpes, Inria, CNRS, Grenoble INP, LJK}
\affil[4]{Courant Institute and Center for Data Science, New York University}
\affil[5]{Laboratoire d’Études Spatiales et d’Instrumentation en Astrophysique, Observatoire de Paris, \newline{} Université PSL, Sorbonne Université, Université Paris Diderot}
\affil[6]{Université Grenoble Alpes, Institut de Planétologie et d’Astrophysique de Grenoble}
}
\begin{document}
\maketitle
\input{sec/abstract}    
\input{sec/introduction}
\input{sec/related_work}
\input{sec/method}
\input{sec/experiments}
\input{sec/conclusion}
    \small
    \bibliographystyle{ieeenat_fullname}
    \bibliography{main}

\input{sec/supplementary}

\end{document}

%% file: sec/abstract.tex
\begin{abstract}
The search for exoplanets is an active field in astronomy, with direct imaging as one of the most challenging methods due to faint exoplanet signals buried within stronger residual starlight. 
Successful detection requires advanced image processing to separate the exoplanet signal from this nuisance component. This paper presents a novel statistical model that captures nuisance fluctuations using a multi-scale approach, leveraging problem symmetries and a joint spectral channel representation grounded in physical principles.
Our model integrates into an interpretable, end-to-end learnable framework for simultaneous exoplanet detection and flux estimation.
The proposed algorithm is evaluated against the state of the art using datasets from the SPHERE instrument operating at the Very Large Telescope (VLT). 
It significantly improves the precision-recall trade-off, notably on challenging datasets that are otherwise unusable by astronomers.
The proposed approach is computationally efficient, robust to varying data quality, and well suited for large-scale observational surveys.
\footnote{Code is available at \url{https://github.com/theobdt/exomild}}
\end{abstract}

%% file: sec/introduction.tex
\vspace*{-5mm}
\section{Introduction}
\label{sec:introduction}

Direct imaging \citep{traub2010direct,follette2023introduction} is an astronomical technique to probe the vicinity of young, nearby stars, where exoplanets and circumstellar disks—structures of dust and gas from which exoplanets can form—are found \citep{keppler2018discovery,haffert2019two}.
Unlike indirect methods \citep{santos2008extra}, which detect exoplanets via secondary effects like gravitational wobbles or transit dimming, direct imaging captures visual evidence of exoplanets and circumstellar disks by recording a \textit{direct image} of their emitted flux. Analyzing this light across spectral bands provides insights into exoplanet properties (e.g., temperature, gravity), atmospheric compositions, molecular abundances, and formation processes \citep{chabrier2000evolutionary,allard2007k,vigan2010photometric}.
\begin{figure}[t!]
	\centering
	\includegraphics[width=.45\textwidth]{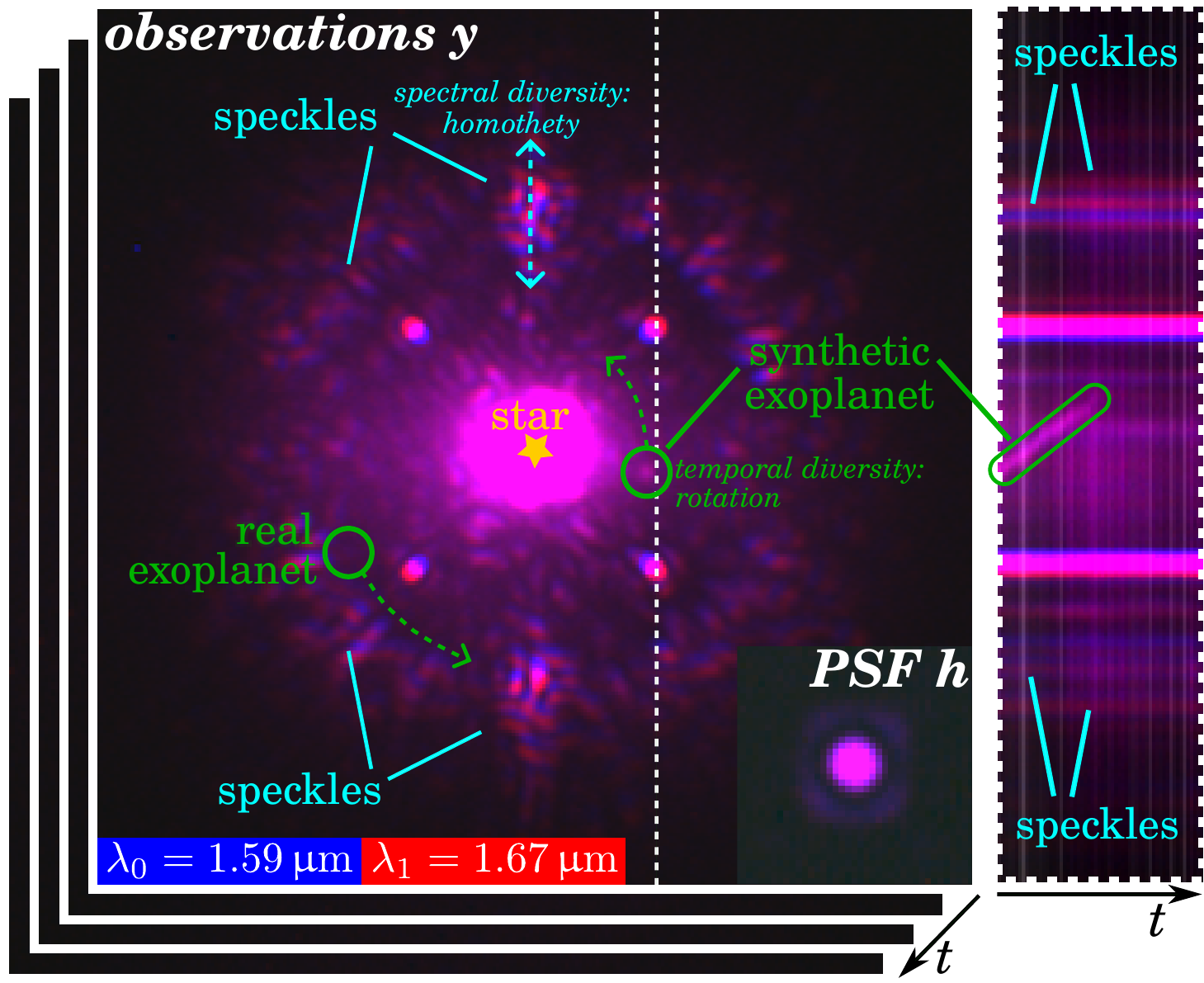}
	\caption{Left: typical observations $\V y$ and PSF $\V h$ from the SPHERE instrument in ASDI mode. The synthetic exoplanet is very bright for illustration purposes. Right: temporal slice along the vertical line.}
	\label{fig:data_asdi}
\end{figure}
While existing technology has enabled imaging of young, giant, gaseous exoplanets, next-generation ground- and space-based telescopes may soon image rocky exoplanets in habitable zones, advancing the search for life beyond our Solar System \citep{chauvin2018direct,currie2022direct}.
Imaging exoplanets is challenging due to the high contrast and angular resolution required. The primary difficulty lies in the large brightness ratio (or \textit{contrast}) between the host star and exoplanets; in the infrared, where gas giant exoplanets’ thermal emission is most detectable, they are typically $10^5$-$10^6$ times fainter.
Additionally, because exoplanets appear close to their host stars from Earth’s perspective, high spatial resolution is essential to separate their faint signals.
To address these challenges, observatories like the VLT are equipped with specialized instruments for direct imaging (e.g., SPHERE \citep{beuzit2019sphere}) and cutting-edge optical technologies. Adaptive optics employs a deformable mirror to obtain sharp images by correcting for atmospheric turbulence in real time \citep{davies2012adaptive}.
Coronagraphs (optical masks) block some starlight, further improving contrast \citep{soummer2004apodized}.

Despite advanced optical devices, direct imaging remains challenging as exoplanet signals are still dominated by a strong \textit{nuisance component} with approximately $10^3$ times higher contrast. This nuisance is mainly composed of structured speckles \citep{fitzgerald2006speckle}, caused by imperfections in optical corrections that allow residual starlight, unblocked by the coronagraph, to leak into images as a spatially correlated diffraction pattern. The speckle pattern is both spatially and spectrally correlated, exhibiting quasi-static behavior across exposures with minor fluctuations over time.
In addition to speckles, other stochastic noise sources--such as thermal background, detector readout, and photon noise--add further contamination. Together, these factors create a non-stationary nuisance that varies in intensity and structure across the field of view, with higher intensity and correlation near the star. This nuisance, which closely resembles the point-like signals expected from exoplanets (instrumental point-spread function off the optical axis), is the primary limitation to direct imaging. Figure \ref{fig:data_asdi} illustrates these characteristics with a dataset recorded by SPHERE.

In this context, dedicated processing is crucial to separate exoplanet signals from the nuisance component \citep{pueyo2018direct}. Observational techniques like angular differential imaging (ADI \citep{marois2006angular}), spectral differential imaging (SDI \citep{sparks2002imaging}), and combined angular and spectral differential imaging (ASDI \citep{vigan2010photometric}) introduce diversity that aids in distinguishing exoplanet signals from the nuisance, see Sec. \ref{sec:direct_models} for resulting image formation models. ADI takes advantage of Earth’s rotation by keeping the telescope pupil fixed, causing off-axis exoplanets to follow a predictable circular trajectory, while the star remains centered. This apparent motion helps separate exoplanet signals from quasi-static speckles. SDI further improves separation by capturing images across multiple spectral channels, where speckles scale quasi-linearly with wavelength while exoplanet signals stay fixed. ASDI combines ADI and SDI to produce a four-dimensional dataset (spatial, temporal, and spectral dimensions).
This diversity in A(S)DI calls for advanced algorithms that can leverage these priors to optimize exoplanet detection and spectral characterization.

We propose a novel hybrid approach that combines the interpretability of a statistical framework with end-to-end learnable components, improving detection sensitivity and characterization accuracy for directly imaged exoplanets. Our main contributions include:
\begin{itemize}
	\item A learnable architecture
	      based on a
	      statistical model integrating pixel correlations across spatial scales and leveraging the spatial symmetries of the nuisance component, improving both detection and characterization.
	\item Statistically reliable detection scores and estimates with uncertainty quantification, essential for astrophysics.
	\item Joint processing of 4-D datasets incorporating the ASDI forward model, shown to significantly boost detection.
\end{itemize}

%% file: sec/related_work.tex
\vspace*{-0.3cm}
\section{Related work}
\label{sec:related_work}

Various methods have been developed to isolate faint planetary signals from stellar nuisance, broadly categorized as subtraction-, statistical-, and learning-based models \citep{pueyo2018direct}.

\vspace*{-0.5cm}
\paragraph{Subtraction-based methods.}

Subtraction-based methods are among the earliest and most common techniques for exoplanet detection in direct imaging, aiming to remove quasi-static speckles that obscure faint signals. The cA(S)DI algorithms \citep{marois2006angular, lagrange2009probable} subtract a reference model by averaging frames and stacking residuals, enhancing the rotating exoplanet signal. TLOCI \citep{marois2014gpi} and its variants (e.g., \cite{wahhaj2015improving,gerard2016planet}) optimize linear combinations of images to model speckles, while KLIP/PCA \citep{soummer2012detection, amara2012pynpoint} employs principal component analysis for low-rank subspace projection. Other approaches, such as non-negative matrix factorization \citep{gonzalez2017vip,ren2018non,ren2020using,pm2023nmf} and LLSG \citep{gonzalez2016low}, decompose data into components to isolate sources, while the RSM algorithm \citep{dahlqvist2020regime,dahlqvist2021auto,dahlqvist2021improving} combines outputs from multiple methods to reduce individual biases. However, these methods often lack statistically rigorous outputs, such as interpretable detection scores and unbiased flux estimates, as they rely on heuristic image combinations rather than fully end-to-end models.

\vspace*{-0.5cm}
\paragraph{Statistical approaches.}
To address the previous limitations, statistical methods model the nuisance using probabilistic frameworks. The ANDROMEDA \citep{cantalloube2015direct} and FMMF \citep{ruffio2017improving} approaches rely on matched filtering, simplifying the nuisance as uncorrelated Gaussian noise. Similarly, SNAP \citep{thompson2021improved} estimates both the nuisance and exoplanet components jointly through maximum likelihood under the same assumption.
The PACO algorithm \citep{flasseur2018exoplanet, flasseur2020paco} improves on this by extending beyond the white noise assumption. It uses a patch-based statistical framework to model the spatial and spectral covariances of the nuisance, effectively capturing the local structure of speckles. This approach draws on techniques from computer vision, such as denoising, restoration, super-resolution, collaborative filtering \citep{dabov2007image}, sparse coding \citep{aharon2006k, mairal2009online}, or mixture models \citep{zoran2011learning, yu2011learning}.

\vspace*{-0.3cm}
\paragraph{Learning-based approaches.}

 Machine learning methods are increasingly used in high-contrast imaging, inspired by advances in fields like photography and biomedical imaging. Early applications in exoplanet detection, such as the SVM model by \cite{fergus2014s4}, leverage structured high-contrast data, while the (NA)-SODINN algorithms \citep{gonzalez2018supervised, cantero2023sodinn} frames detection as a binary classification problem, using KLIP/PCA-processed patches with random forests or CNNs. Despite their effectiveness, SODINN struggles with high false alarm rates and complex hyperparameter tuning \citep{cantalloube2020exoplanet}. Generative adversarial networks (GANs) have also been applied to simulate nuisance patterns for training deep learning models \citep{yip2020pushing}. Approaches like TRAP \citep{samland2021trap} and HSR \citep{gebhard2022half} handle unmixing via regularized regression, modeling nuisance evolution with signal-free reference data. Deep PACO \citep{flasseur2023combining, flasseur2024deep} combines statistical modeling with deep learning, leveraging PACO's nuisance statistical model and a CNN to detect exoplanets and refine residual mismatches.

All these algorithms are observation-dependent, building nuisance models directly from the dataset of interest. This dependency hinders detection near the host star due to (i) significant temporal fluctuations in the nuisance and (ii) residual self-subtraction, where part of the exoplanet signal is mistakenly removed. Recent observation-independent approaches address these limitations by using archival survey data to model the nuisance. Super-RDI \citep{sanghi2024efficiently} extends KLIP/PCA for large observational databases, while ConStruct \cite{wolf2024direct} uses an auto-encoder to learn typical speckle patterns, and \citep{chintarungruangchai2023possible} employs a discriminative nuisance model. 
MODEL\&CO \citep{bodrito2024model} improves deep PACO with a deep statistically-modeled nuisance framework. 
Our proposed approach, ExoMILD (Exoplanet imaging by MIxture of Learnable Distributions), belongs to this new category.

%% file: sec/method.tex
\section{Image formation models}
\label{sec:direct_models}

In direct imaging, a stellar system is observed through an optical system that includes the atmosphere, telescope, and scientific instrument. The response of this system to a point source (e.g., an exoplanet) is defined by its point-spread function off the optical axis (off-axis PSF), describing how light from the source is distributed across the sensor.
However, as noted in Sec.~\ref{sec:introduction}, residual aberrations uncorrected by adaptive optics produce a quasi-static, structured speckle pattern.
To mitigate speckles through numerical processing, several observational strategies are used, as detailed next.

\subsection{Angular Differential Imaging (ADI)}
\label{subsec:adi}

In ADI, the field derotator of the telescope is turned off
causing the field of view (including any exoplanets) to rotate around the target star due to Earth's rotation. Meanwhile, the optical system remains stationary, causing the speckles to remain fixed. This distinction
helps separating exoplanet signals from speckle noise.
Formally, let $\V y$ in $\mathbb{R}^{T \times H \times W}$ denote the sequence of measurements from the observation, where $T$ is the number of exposures and $H$, $W$ represent the pixel dimensions of each exposure. Each frame $\V y_t$ in $\mathbb{R}^{H \times W}$ can then be represented as:
\begin{equation}
	\V y_t = {\V s}_t + \sum\nolimits_{k=0}^{K-1} \alpha^{(k)} \, \V h(\V x^{(k)}_t) + \V \epsilon_t \,,
	\label{eq:adi_direct_model}
\end{equation}
where $\V s_t$ in $\mathbb{R}^{H \times W}$ is the speckles component,
$K$ is the (unknown) number of exoplanets,
$\V x^{(k)}_t$ is the position of exoplanet $k$ at time $t$,
$\V h(\V x)$ in $\mathbb{R}^{H \times W} $ is the PSF model centered on position $\V x$ in $\mathbb{R}^2$,
$\alpha^{(k)}$ in $\mathbb{R}_+$ is the flux of the exoplanet,
and $\V \epsilon_t$ in $\mathbb{R}^{H \times W}$ is stochastic noise.
The position of exoplanet $k$ at time $t$ can be written as $\V x^{(k)}_t = r \big( \V x^{(k)}_0, \phi_t \big)$,
where $\V x^{(k)}_0$ in $\mathbb{R}^{2}$ is its initial position on frame $\V y_0$,
$\phi_t$ in $\mathbb{R}$ is the predictable \textit{parallactic angle} at time $t$, defined as the cumulated apparent rotation (induced by Earth's rotation) since the start of the sequence,
and $r: \mathbb{R}^2\times\mathbb{R} \to \mathbb{R}^2$ defines a rotation trajectory centered on the star.

\subsection{Angular Spectral Differential Imaging (ASDI)}
\label{subsec:asdi}

The speckles pattern induced by the star can be understood as the PSF on the optical axis (on-axis PSF) of the optical system.
According to diffraction laws, this pattern scales homothetically with wavelength, creating a chromatic effect similar to chromatic aberrations in photography. In ASDI, this spectral scaling helps further disentangle exoplanet signals from speckles.
In this case, each observation is denoted as $\V y$ in $\mathbb{R}^{C\times T \times H \times W}$, where $C$ is the number of channels.
The forward model \eqref{eq:adi_direct_model} becomes
\begin{equation}
	\V y_{c, t} = \beta_c \, {\M D}_{\lambda_c / \lambda_0} \ {\V s}_{0, t} + \sum\nolimits_{k=0}^{K-1} \alpha^{(k)}_c \V h_c(\V x^{(k)}_t) + \V \epsilon_{c, t} \,,
	\label{eq:asdi_direct_model}
\end{equation}
\noindent where $\beta_c$ in $\mathbb{R}$ is the amplitude of the speckles in channel $c$,
${\M D}_{\lambda_c / \lambda_0}$ is the homothety operator aligning channel $0$ on channel $c$ with a dilation coefficient defined as the ratio of their wavelengths,
$\V s_{0, t}$ in $\mathbb{R}^{H \times W}$ is speckles pattern at wavelength $\lambda_0$,
$\V \epsilon_{c, t}$ in $\mathbb{R}^{H \times W}$ is the additive thermal noise.
In this setting, the exoplanet flux $\alpha_c$ and the off-axis PSF $\V h_c$ now depend on channel $c$.

\section{Proposed method}
\label{sec:proposed_method}

\begin{figure*}[t]
	\includegraphics[width=\textwidth]{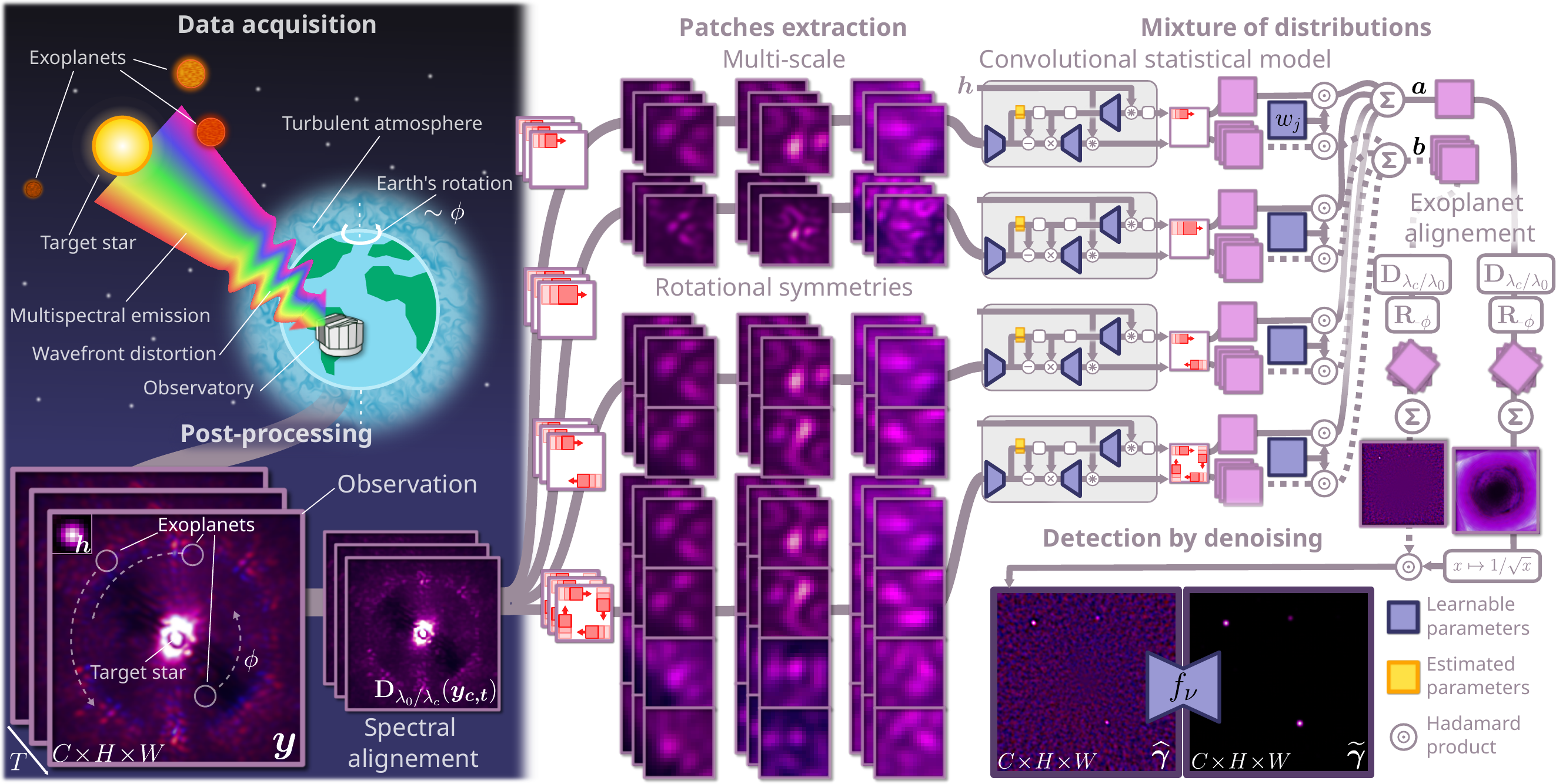}
	\caption{
		Workflow of the proposed method: it exploits both the spectral behavior of speckles and the apparent motion of exoplanets to disentangle the exoplanet signal from the nuisance component in the observations $\V y$.
		To achieve this, local patches of the nuisance are modeled as Gaussian distributions, leveraging problem symmetries and incorporating multiple scales.
		These patches are fed to our convolutional statistical model, and combined to form a detection map.
		Additionally, a learned object prior, represented by a UNet $f_\nu$, is introduced to denoise this detection map produced by the statistical model. This approach results in an end-to-end learnable architecture.
	}

\end{figure*}

\subsection{Convolutional statistical model}
\label{subsec:statistical_model}

\begin{figure}[t]
	\includegraphics[width=.47\textwidth]{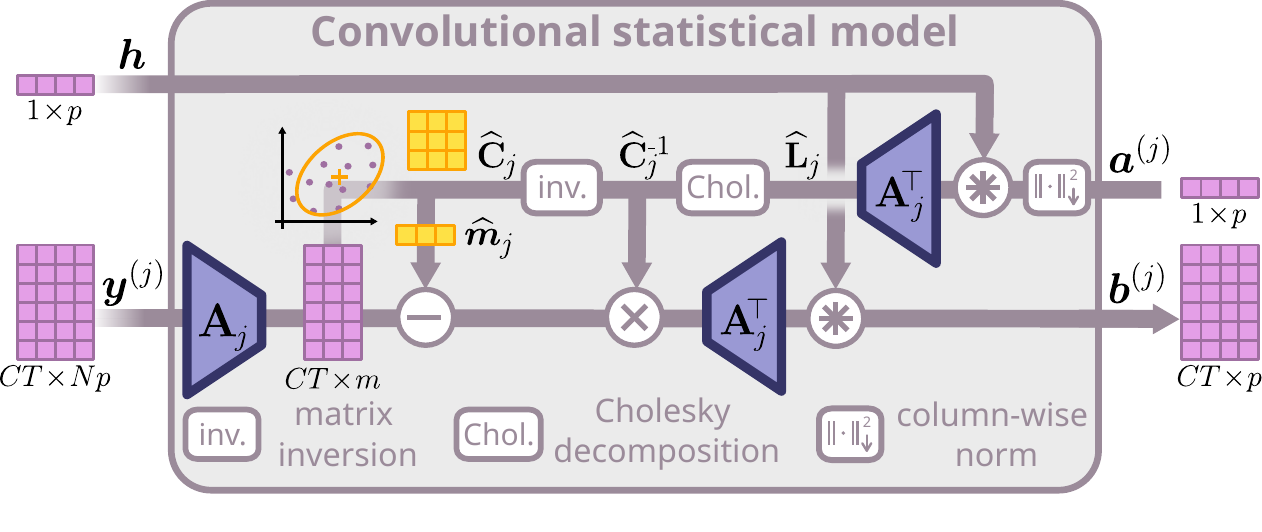}
	\caption{
		Proposed convolutional statistical model: spectrally aligned speckles patches, indexed by $j$, with dimensions $Np$ and $CT$ samples, are first linearly projected into a lower-dimensional space of size $m$.
		In this space, the parameters of the Gaussian distribution $\widehat{\V m}_j$ and $\widehat{\M C}_j$ are estimated and subsequently combined with the PSF $\V h$ to compute the terms $\V a^{(j)}$ and $\V b^{(j)}$. As detailed in Appendix~\ref{sec:conv_ab}, the efficient computation of $\V a^{(j)}$ relies on the Cholesky decomposition of the precision matrix $\widehat{\M C}^{-1}_j$.
	}
\end{figure}

\paragraph{Local Gaussian model of speckles.}
Building on the PACO algorithm \cite{flasseur2018exoplanet,flasseur2020paco}, we propose to capture local spatial correlations between pixels of the nuisance term.
We denote by $\mathcal{G}_p$ the grid of spatial pixels, such that $|\mathcal{G}_p| = HW$
and $\forall i$ in $\llbracket 0, HW - 1 \rrbracket, \V x_i$ in $\mathbb{R}^2$.
Then, we model the statistical distribution of collections of patches positioned on a spatial grid $\mathcal{G}_d$,
such that $\mathcal{G}_d \subset \mathcal{G}_p$, with $M \vcentcolon= | \mathcal{G}_d|$.
We denote by $\V y_t^{(j)}$ in $\mathbb{R}^{p}, \ \forall (j, t) \in \llbracket 0, M - 1 \rrbracket \times \llbracket 0, T - 1 \rrbracket$ the observed patch in spatial location $j$ at time $t$.
In absence of exoplanet, i.e., when only the nuisance component is present, each collection of patches $\{\V y_t^{(j)} \}_{t=0:T-1}$ is modeled by a multivariate Gaussian:
\begin{equation}
	\forall (j, t), \ \V y_t^{(j)} \sim \mathcal{N}(\V m_j, \M C_j) \,,
\end{equation}
\noindent with $\V m_j$ in $\mathbb{R}^p$ and $\M C_j$ in $\mathbb{R}^{p \times p}$ the mean and covariance of the Gaussian distribution.
These parameters are estimated by maximum likelihood and shrinkage of the covariance matrix.
They are denoted as $\widehat{\V m}_j$ and $\widehat{\M C}_j$ in the following.
Additional details are presented in Appendix~\ref{sec:estimation_params}.

\vspace*{-0.3cm}
\paragraph{Detection criterion at a given position.}
This local statistical model estimates the time-invariant flux $\widehat{\alpha}$ of an exoplanet at position $\V x_0$ in $\mathbb{R}^2$ in the first frame by maximizing the global likelihood:
\begin{equation}
	\widehat{\alpha} = \argmax_{\alpha} \ell(\alpha, \V x_0) \,.
\end{equation}
Assuming that collections of patches on the trajectory of an exoplanet are independent,
the likelihood becomes:
\begin{equation}
	\ell(\alpha, \V x_0) = \prod\nolimits_t \mathbb{P} \left( \V y_t^{(i_t)} - \alpha \V h^{(i_t)}(\V x_t) \, | \,  \widehat{\V m}_{i_t}, \widehat{\M C}_{i_t} \right)\,,
	\label{eq:ml_paco}
\end{equation}
\noindent where $\V x_t = r(\V x_0, \phi_t)$ is the predictive position of the exoplanet at time $t$,
$i_t$ the index of the patch centered on position $\lfloor \V x_t \rceil$ in $\mathcal{G}_p$,
and $\V h^{(i_t)}(\V x_t)$ in $\mathbb{R}^p$ the patch $i_t$ of the off-axis PSF model centered on $\V x_t$. We propose to consider the modified likelihood:
\begin{equation}
	\ell(\alpha, \V x_0) = \prod_t \prod_{j \in S(\V x_t)}\mathbb{P} \left( \V y_t^{(j)} - \alpha \V h^{(j)}(\V x_t) \, | \, \widehat{\V m}_{j}, \widehat{\M C}_{j} \right)^{w_j}\,,
	\label{eq:likelihood_pacon}
\end{equation}
where $S(\V x_t)$ represents the subset of modeled distributions for patches around location $\V x_t$, with each distribution $j$ weighted by $w_j$ in $\mathbb{R}^{+}$, where $\sum w_j = 1$. Compared to \eqref{eq:ml_paco}, this approach is more robust, as it aggregates overlapping patch contributions at each time-step, modeling the nuisance component as a convolutional process influenced by multiple, overlapping noise sources. It is also more computationally efficient, as it does not require a dense grid $\mathcal{G}_d$ for patch collections, thereby reducing the parameter estimation load. This efficiency is essential for incorporating the model into an end-to-end learning framework (see Sec.~\ref{subsec:learning}). Finally, this flexible formulation can accommodate additional correlation types (see Sec. \ref{subsec:mixture_distributions}).
Solving problem~\eqref{eq:likelihood_pacon} leads to the flux estimator $\widehat{\alpha}$ and its standard deviation $\widehat{\sigma}_{\alpha}$:
\begin{equation}
	\widehat{\alpha} = \frac{\sum_t b_t(\V x_t)}{ \sum_t a_t(\V x_t)} \,, \quad \widehat{\sigma}_\alpha = \frac{1}{\sqrt{\sum_t a( \V x_t )}}\,,
	\label{eq:flux_estimator}
\end{equation}
\noindent where
\begin{align}
	b_t(\V x_t) & = \sum_{j \in S(\V x_t)} w_j \, \V h^{(j)}(\V x_t)^\top \,  \widehat{\M C}_{j}^{-1} \, ( \V y_t^{(j)} - \widehat{\V m}_{j}) \,, \label{eq:b} \\
	a(\V x_t)   & = \sum_{j \in S(\V x_t)} w_j \, \V h^{(j)}(\V x_t)^\top \, \widehat{\M C}_{j}^{-1} \, \V h^{(j)}(\V x_t) \,. \label{eq:a}
\end{align}
To assess the probability of an exoplanet at $\V x_0$, we use the generalized likelihood ratio test (GLRT) to statistically evaluate the parameter $\alpha$. Under the null hypothesis $\mathcal{H}_0$ where $\alpha = 0$ (indicating no exoplanet), the statistics
\begin{equation} \label{eq:detection_score}
	\widehat{\gamma} = \widehat{\alpha} \big/ {\widehat{\sigma}_\alpha}
	= \Big( \sum\nolimits_t b_t(\V x_t) \Big) \Big/ \Big( \sqrt{\sum\nolimits_t a(\V x_t) } \Big) \,,
\end{equation}
is controlled and follows a Gaussian distribution $\mathcal{N}(0, 1)$.
When evaluating the probability of presence of an exoplanet, we test against the alternative hypothesis $\mathcal{H}_1$: $\alpha > 0$.
The statistics $\widehat{\gamma}$ can be directly mapped to a probability of detection.
It can also be interpreted as the output signal-to-noise ratio through the statistical model,
and in practice, this is the detection score used by astronomers \citep{thiebaut2016fast}.
To obtain the dense counterparts $\widehat{\V \gamma}, \widehat{\V \alpha}, \widehat{\V \sigma}$ in $\mathbb{R}^{H \times W}$, we adopt a fast approximation similar to \cite{flasseur2018exoplanet} detailed in Appendix~\ref{sec:dense_map}.

\vspace*{-0.3cm}
\paragraph{Iterative procedure for characterization.}
Astrometry is the task of precisely estimating the flux $\alpha$ and the sub-pixel position $\V x_0$ of an exoplanet.
This can be achieved by jointly optimizing the likelihood defined in Eq.~\eqref{eq:likelihood_pacon}.
Denoting by $\V z = [\alpha, \V x_0]$ in $\mathbb{R}^3$, the gradient $\V g$ and Hessian matrix $\M H$ of the neg-log-likelihood are decomposed as follows:
\begin{equation}
	\V g(\V z) = \sum_t \! \sum_{j \in S(\V x_t)} \! w_j \V g_j(\V z),\,
	\M H(\V z) = \sum_t \! \sum_{j \in S(\V x_t)} \! w_j \M H_j(\V z)
	\nonumber
\end{equation}
\noindent where
$\V g_j$ and $\M H_j$ represent the gradient and Hessian operators for each distribution $j$, respectively. These operators are responsible for updating the statistical parameters $\V m_j$ and $\M C_j$, which are initially biased due to the presence of the exoplanet.
Further details are provided in Sec.~\ref{sec:iteration_astrometry}.
Each iteration indexed by $l$ writes as:
\begin{equation}
	\V z^{(l+1)} = \V z^{(l)} - \M H(\V z^{(l)})^{-1} \V g(\V z^{(l)}) \,.
\end{equation}

\subsection{Extensions of the statistical model: mixture of distributions}

\label{subsec:mixture_distributions}

We now build on the convolutional statistical model presented in Sec. \ref{subsec:statistical_model} to introduce a multi-scale statistical model of the nuisance component.
\vspace*{-0.3cm}
\paragraph{Linear projection.}
Instead of modeling correlations between the pixels of a patch,
we propose to model correlations between projected linear features:
\begin{equation}
	\forall (j, t), \quad \M A \, \V y^{(j)}_t \sim \mathcal{N}(\V m_j, \M C_j) \,,
\end{equation}
\noindent where $\M A$ in $\mathbb{R}^{m \times p}$ is the projection matrix and $m \le p$.
This is a generalization of the model presented in Sec.~\ref{subsec:statistical_model}, for which $\M A = \M I_p$.
The terms $\V b_t$ and $\V a$ introduced in Eqs.~\eqref{eq:b}-\eqref{eq:a} become:
\begin{align} \label{eq:bl}
	b_{i, t} & = \sum_{j \in S(\V x_i)} w_j \, \V h_{ji}^\top \, \M A_j^\top \, \widehat{\M C}_{j}^{-1} \ \big(\M A_j \, \V y_t^{(j)} - \widehat{\V m}_j \big) \,, \\
	a_i      & = \sum_{j \in S(\V x_i)} w_j \, \V h_{ji}^\top \M A_j^\top \, \widehat{\M C}_{j}^{-1} \, \M A_j \, \V h_{ji} \,. \label{eq:al}
\end{align}
The learnable projection $\M A$ decorrelates the feature space, where the statistical distribution is defined, from the pixel space, enabling effective handling of long-range correlations.
\vspace*{-0.3cm}
\paragraph{Multi-scale approach.}

So far, we have shown how our model can combine multiple neighboring local distributions through weighted averaging of the terms $\V b_t$ and $\V a$.
Extending this approach, we propose to integrate distributions across multiple spatial scales, by varying the patch sizes $p$.
We denote $\mathcal{P}$ the set of patch sizes chosen.
We expand the set $S(\V x_t)$ of modeled distributions to include different scales, such that:
\begin{equation}
	S(\V x_t) =\bigcup\nolimits_{p \in \mathcal{P}} S_p(\V x_t) \,,
\end{equation}
where $S_p(\V x_t)$ is the subset of patches with patch size $p$ containing $\V x_t$.
The linear projection is particularly useful in controlling the dimensionality of the feature space for larger patches by setting $m < p$, thereby maintaining computational efficiency.

\vspace*{-0.3cm}
\paragraph{Leveraging rotational symmetries.}
In direct imaging, the speckle pattern exhibits approximate central and rotational symmetries, see Fig.~\ref{fig:data_asdi}.
These symmetries arise from near-isotropic distortions of the wavefront and from the optical system's near-circular symmetry. Symmetries of the speckles can be theoretically understood through a series expansion of the diffraction pattern, using increasing powers of the Fourier transform of the residual phase error \citep{perrin2003structure,soummer2007speckle,ribak2008fainter}.
Leveraging these symmetries is essential for constructing a more robust model of the speckles, particularly to mitigate the effects of \textit{self-subtraction}. This effect is a well-known challenge in direct imaging that arises when limited parallactic rotation results in the contamination of speckle parameters by the exoplanet signal, ultimately reducing detection sensitivity \citep{milli2012impact}.
To incorporate these symmetries, we propose modeling the joint distribution of patches extracted from the same location after rotating the observation by $2 \pi / N$,
also known as \textit{$N$-fold rotational symmetry}:
\begin{equation}
	\forall (j, t), \quad \M A \big[  {\M R}_{2 \pi n/ N} (\V y_t)^{(j)} \big]_{n=0:N-1} \sim \mathcal{N}(\V m_j, \M C_j) \,,
\end{equation}
\noindent where $\M A$ is in  $\mathbb{R}^{m \times Np}$.
The parameters of this joint distribution are less subject to self-subtraction as it is very unlikely to observe exoplanets simultaneously in all jointly modeled patches.
We use a mixture model with $N=1, 2, 4$.
\vspace*{-0.3cm}
\paragraph{Joint multi-spectral modeling.}
In ASDI, the spectral channels are related by a homothety, as given in Eq.~\eqref{eq:asdi_direct_model}.
We propose to leverage this relationship, and model:
\begin{equation}
	\forall (j, c, t), \quad
	\M A \, {\beta_{c, j}^{-1}} \, {\M D}_{\lambda_0 / \lambda_c}(\V y_{c,t})^{(j)} \sim \mathcal{N}(\V m_j, \M C_j) \,.
\end{equation}
In this setting, the estimators of statistical parameters $\widehat{\V m}_j$ and
$\widehat{\M C}_j$ are less prone to self-subtraction,
as the spectral diversity reduces the ambiguity between speckles and exoplanet signals.
The estimator of the local amplitude $\widehat{\beta}_{c, j}$
is computed as the standard deviation of pixel values across $\{{\M D}_{\lambda_0 / \lambda_c}(\V y_{c,t})^{(j)} \}_{t}$.

\subsection{End-to-end trainable approach}
\label{subsec:learning}

\paragraph{Problem statement.}
We propose to combine our statistical model on the nuisance component with a learnable prior on the exoplanet signals.
This approach can be formulated as an optimization problem:
\begin{equation} \label{eq:pb_learned}
	\widehat{\V \alpha}  = \argmin_{\V \alpha \in \mathbb{R}^{H \times W}} \varphi_\theta(\V \alpha, \chi(\V y)) + \psi_\nu(\V \alpha)\,,
\end{equation}
where $\varphi$ is the data-fitting term, corresponding to the negative log-likelihood of our statistical model of the nuisance component described in Sec.~\ref{subsec:statistical_model},
and $\psi$ is a prior on exoplanet signals.
We denote by $\theta, \nu$ the learnable parameters associated with these terms,
and by $\chi(\V y)$ the parameters estimated for each observation.
In practice $\chi(\V y) = \{\widehat{\V m}_j, \widehat{\M C}_j, \widehat{\beta}_{c, j} \}$, $\theta = \{\M A_j, w_j\}_j$, and $\nu$
corresponds to the parameters of the neural network implementing $\psi$.
\vspace*{-0.3cm}
\paragraph{Detection by denoising.}
We propose a two-step approach to obtain the detection score corresponding to Eq.~\eqref{eq:pb_learned}.
First, we compute the detection score corresponding to the statistical model $\varphi$, which admits a fast approximation denoted by $\V \gamma \in \mathbb{R}^{H \times W}$ in the following, and given by Eq.~\eqref{eq:detection_score_fast}.
We recall that under the null hypothesis, i.e., when no exoplanet is present, the elements of $\V \gamma$ follow a Gaussian distribution $\mathcal{N}(0, 1)$.
Therefore, extracting the signals of exoplanets corresponds to denoising $\V \gamma$ to remove this background noise.
We propose to achieve this step using a neural network, such that:
\begin{equation} \label{eq:denoised_score}
	\widetilde{\V \gamma} = f_\nu \left( \widehat{\V \gamma} \right) \,,
\end{equation}
where $f_\nu$ is the denoiser implemented by the neural network, and $\widetilde{\V \gamma}$ the final detection score.
\vspace*{-0.3cm}
\paragraph{Training objective.}
We suppose that the denoised detection score can be decomposed similarly to the GLRT form provided in Eq.~\eqref{eq:detection_score} for the statistical model.
This leads to $\widetilde{\V \gamma} = \widetilde{\V \alpha} / \widetilde{\V \sigma}_e$,
where $\widetilde{\V \alpha} ,  \widetilde{\V \sigma}$ in $\mathbb{R}^{H \times W}$ are
the estimated flux of the exoplanet for each pixel, and the standard deviation denoting the uncertainty associated with it.
Additionally, we assume that the uncertainty $ \widetilde{\V \sigma}$ remains unaffected by the neural network, as it is primarily driven by the high variability of the speckles already captured by the statistical model.
Consequently, both components can be expressed as:
\begin{equation}
	\widetilde{\V \alpha} = f_\nu ( \widehat{\V \alpha} / \widehat{\V \sigma} ) \times \widehat{\V \sigma},\,\quad
	\widetilde{\V \sigma} = \widehat{\V \sigma} \,.
\end{equation}
This formulation yields a pixel-wise Gaussian distribution: $\mathcal{N}(\widetilde{\V \alpha},\widetilde{\V \sigma})$.
The learnable parameters $\theta$ and $\nu$ of our model are optimized by minimizing the negative log-likelihood between estimates $(\widetilde{\V \alpha},\, \widetilde{\V \sigma})$ and ground truth $\V \alpha_{\text{gt}}$:
\begin{equation}
	\mathcal{L}(\widetilde{\V \alpha}, \widetilde{\V \sigma}, \V \alpha_{\text{gt}})
	= 0.5 \, (\widetilde{\V \alpha} - \V \alpha_{\text{gt}})^2 / \widetilde{\V \sigma}^2
	+ \log \widetilde{\V \sigma}\,.
\end{equation}

\vspace*{-0.3cm}
\paragraph{Model ensembling.}
To improve robustness, we combine outputs from multiple trained models.
Specifically, given outputs $\widetilde{\V \alpha}$ and $\widetilde{\V \sigma}$,
we can define $\widetilde{\V a}$ and $\widetilde{\V b}$ such that $\widetilde{\V \sigma} = 1/ \sqrt{\widetilde{\V a}}$ and $\widetilde{\V \alpha} = \widetilde{\V b} / \widetilde{\V a}$.
For $Q$ models indexed by $q$, the outputs are aggregated as follows:
\begin{equation}
	\widetilde{\V a}_Q = \sum\nolimits_q \widetilde{\V a}_q / Q, \quad
	\widetilde{\V b}_Q = \sum\nolimits_q \widetilde{\V b}_q / Q  \,.
\end{equation}
The combined detection score, $\widetilde{\V \gamma}_Q = \widetilde{\V b}_Q / \sqrt{\widetilde{\V a}_Q}$, represents a weighted average of the individual detection scores.

\vspace*{-0.3cm}
\paragraph{Calibration.}
The trained model needs to be calibrated in order to relate the detection score to a probability of false alarm.
We follow the procedure outlined in \cite{bodrito2024model}, and rely on a separate calibration dataset to estimate the cumulative distribution function of $\widetilde{\V \gamma}$ under the null hypothesis. Additional details are provided in the Supplementary Material.

%% file: sec/experiments.tex
\section{Experiments}
\label{sec:experiments}

\subsection{Data, algorithms and evaluation protocol}
\label{subsec:data_algorithms}

\begin{table}[t!]
	\begin{center}
		\label{tab:table1}
		\resizebox{\columnwidth}{!}{%
			\begin{tabular}{@{}l|l|ccc}
				\toprule
				                                                    & Spatial scales & N=1               & N=1,2             & N=1, 2, 4                  \\
				\midrule
				\multirow{4}{*}{\begin{sideways}ADI\end{sideways}}  & $8\times8$     & $0.554 \pm 0.005$ & $0.571 \pm 0.005$ & $0.575 \pm 0.005$          \\
				                                                    & +$16\times16$  & $0.561 \pm 0.005$ & $0.573 \pm 0.004$ & $0.579 \pm 0.005$          \\
				                                                    & +$32\times32$  & $0.567 \pm 0.005$ & $0.575 \pm 0.005$ & $0.580 \pm 0.005$          \\
				                                                    & +$64\times64$  & $0.569 \pm 0.005$ & $0.577 \pm 0.005$ & $\mathbf{0.581} \pm 0.005$ \\
				\midrule
				\multirow{4}{*}{\begin{sideways}ASDI\end{sideways}} & $8\times8$     & $0.713 \pm 0.005$ & $0.719 \pm 0.005$ & $0.723 \pm 0.005$          \\
				                                                    & +$16\times16$  & $0.720 \pm 0.005$ & $0.723 \pm 0.005$ & $0.725 \pm 0.005$          \\
				                                                    & +$32\times32$  & $0.720 \pm 0.005$ & $0.725 \pm 0.005$ & $\mathbf{0.726} \pm 0.005$ \\
				                                                    & +$64\times64$  & $0.720 \pm 0.005$ & $0.725 \pm 0.004$ & $\mathbf{0.726} \pm 0.004$ \\
				\bottomrule
			\end{tabular}
		}
		\caption{
			Impact of multi-scale and $N$-fold rotational symmetries on detection performance (AUC) of our statistical model.
		}
		\label{table:ablations}
	\end{center}
\end{table}
\paragraph{Datasets.}
We use SPHERE datasets, a cutting-edge exoplanet finder instrument at the VLT \citep{beuzit2019sphere}. Raw observations were sourced from the public data archive of the European Southern Observatory
and calibrated with public tools \citep{pavlov2008sphere,delorme2017sphere} from the
High-Contrast Data Center.
This results in science-ready 4-D datasets $\V y$ (with $L=2$ spectral channels, $T \in \llbracket 15, 300\rrbracket$ temporal frames, and $H \times W = 1024^2$ pixels per image), along with the off-axis PSF $\V h$, wavelengths $\lambda_c$, and parallactic angles $\phi_t$ for algorithm input.
Many algorithms have achieved optimal detection sensitivity far from the star, limited by photon noise \citep{flasseur2024deep, bodrito2024model}. Thus, our analysis focuses on a smaller, star-centered region of $H = W = 256$ pixels, where detection sensitivity can still be significantly improved \citep{flasseur2024deep,bodrito2024model}.
\vspace*{-0.4cm}
\paragraph{Training procedure.}
For training, we use 220 observations from the SHINE-F150 large survey of SPHERE \citep{langlois2021sphere}. For testing, we select 8 datasets representing typical diversity in observing conditions and instrumental settings (e.g., parallactic rotation amplitude $\Delta_{\phi} = \left| \phi_{T-1} - \phi_0 \right|$).
Five test datasets (on stars HD 159911, HD 216803, HD 206860, HD 188228, HD 102647) are used for benchmarking against state of the art methods and conducting a model ablation analysis.
To evaluate performance, we simulate synthetic faint point-like sources mimicking exoplanet signatures and inject them into real data. This simulation procedure is common practice in direct imaging to ground the actual performance of algorithms because it is very realistic \citep{gonzalez2018supervised,cantalloube2020exoplanet,dahlqvist2020regime,daglayan2022likelihood,cantero2023sodinn,flasseur2024deep}. Any real source indeed takes the form of the off-axis PSF, which is measured immediately before and after the main observation sequence by offsetting the star from the coronagraph. This simulation procedure is essential due to the lack of ground truth and the limited number of exoplanets (only a few dozen) detected by direct imaging to date.
In addition, we rely on simulated sources for training our deep model, which is also a standard practice for learning-based techniques.
Three datasets of star HR 8799, hosting three known exoplanets in the field of view \citep{marois2008direct,marois2010images}, are also used as full real data.
\vspace*{-0.4cm}
\paragraph{Baselines.}
For detection, we benchmark the proposed algorithms against methods from different classes described in Sect.~\ref{sec:related_work}. Selection criteria are (i) code availability (often limited in direct imaging), (ii) relevance, and (iii) widespread use.
In this context, we include the cADI \citep{marois2006angular}, KLIP/PCA \citep{soummer2012detection} subtraction-based methods, as these are implemented in most data processing pipelines \citep{delorme2017sphere,stolker2019pynpoint,christiaens2023vip}, have been responsible for detecting nearly all imaged exoplanets --including the most recent discoveries \citep{wagner2023direct, mesa2023af}-- and remain heavily utilized by astronomers.
Concerning statistical methods, we focus on PACO \citep{flasseur2018exoplanet,flasseur2020paco} that uniquely accounts for data correlations.
PACO has consistently outperformed KLIP/PCA in large observational surveys \citep{chomez2023preparation,delorme2024giant} and achieved state of the art performance on SPHERE data in a community benchmark, surpassing various subtraction-based, statistical, and learning-based approaches \citep{cantalloube2020exoplanet}.
We also evaluate the proposed approach against MODEL\&CO \citep{bodrito2024model}, a recent hybrid method that shown to outperform cADI, KLIP/PCA, and (deep) PACO \citep{bodrito2024model}. For cADI and KLIP/PCA, we use the VIP Python package \citep{gonzalez2017vip,christiaens2023vip}, fine-tuning parameters (e.g., PCA modes) to optimize detection scores guided by the ground truth. PACO and MODEL\&CO were processed by their authors using data-driven settings. For flux estimation, we compare only with PACO due to computational constraints.
We conduct all analyses in both ADI and ASDI modes (where supported) to assess the gains from joint spectral processing.
\begin{table}[t!]
	\begin{center}
		\resizebox{0.8\columnwidth}{!}{%
			\begin{tabular}{ccc}
				\toprule
				Method   & flux error (ARE) & position error (RMSE) \\
				\midrule
				PACO     & $ 0.56$          & $0.21$                \\
				Proposed & $\mathbf{0.51}$  & $\mathbf{0.11}$       \\
				\bottomrule
			\end{tabular}
		}
		\caption{Comparison of statistical models for flux estimation.}
		\label{table:charac}
	\end{center}
\end{table}
\vspace*{-0.3cm}
\paragraph{Metrics.}
The detection metric used is the area under the receiver operating characteristic curve (AUC), representing the true positive rate against the false discovery rate obtained by varying the detection threshold. Higher AUC values indicate better performance. This standard metric in direct imaging \citep{gonzalez2016low,flasseur2018exoplanet,cantalloube2020exoplanet} captures the precision-recall trade-off and allows fair algorithm comparisons, as a common threshold does not ensure consistent false alarm rates due to the lack of statistical grounding in some detection maps, see Sec.~\ref{sec:introduction}.
The primary metric is the absolute relative error (ARE) between the ground truth and estimated flux, with lower values indicating better performance. We also report the root mean square error (RMSE) for sub-pixel localization of exoplanets.

\begin{table*}[t!]
	\begin{center}
		\label{tab:table1}
		\resizebox{1.95\columnwidth}{!}{%
			\begin{tabular}{cl|cccccc}
				\toprule 
				\textbf{Modality} & \textbf{Method} & HD159911 (54\degree)       & HD216803 (23\degree)       & HD206860 (11\degree)       & HD188228 (6\degree)        & HD102647 (2\degree)        & average AUC                \\
				\midrule
				\multirow{5}{*}{ADI}
				                  & cADI            & $0.288 \pm 0.014$          & $0.422 \pm 0.007$          & $0.489 \pm 0.010$          & $0.303 \pm 0.014$          & $0.343 \pm 0.009$          & $0.369 \pm 0.005$          \\
				                  & PCA             & $0.634 \pm 0.010$          & $0.643 \pm 0.011$          & $0.505 \pm 0.009$          & $0.392 \pm 0.011$          & $0.218 \pm 0.011$          & $0.478 \pm 0.005$          \\
				                  & PACO            & $0.629 \pm 0.006$          & $0.669 \pm 0.012$          & $0.579 \pm 0.015$          & $0.517 \pm 0.015$          & $0.207 \pm 0.012$          & $0.520 \pm 0.006$          \\
				                  & MODEL\&CO       & $0.653 \pm 0.010$          & $0.731 \pm 0.011$          & $0.646 \pm 0.012$          & $\textbf{0.638} \pm 0.014$ & $\textbf{0.554} \pm 0.009$ & $\mathbf{0.645} \pm 0.005$ \\
				                  & Proposed        & $\mathbf{0.673} \pm 0.009$ & $\mathbf{0.740} \pm 0.010$ & $\textbf{0.661} \pm 0.012$ & $0.631 \pm 0.013$          & $0.518 \pm 0.012$          & $\mathbf{0.645} \pm 0.005$ \\
				\midrule
				\multirow{4}{*}{ASDI}
				                  & cASDI           & $0.448 \pm 0.009$          & $0.537 \pm 0.013$          & $0.451 \pm 0.007$          & $0.352 \pm 0.017$          & $0.294 \pm 0.012$          & $0.417 \pm 0.006$          \\
				                  & PCA             & $0.694 \pm 0.013$          & $0.696 \pm 0.007$          & $0.552 \pm 0.011$          & $0.398 \pm 0.011$          & $0.236 \pm 0.009$          & $0.515 \pm 0.005$          \\
				                  & PACO            & $0.698 \pm 0.015$          & $0.768 \pm 0.008$          & $0.710 \pm 0.014$          & $0.700 \pm 0.009$          & $0.589 \pm 0.014$          & $0.693 \pm 0.005$          \\
				                  & Proposed        & $\textbf{0.731} \pm 0.009$ & $\textbf{0.804} \pm 0.013$ & $\textbf{0.747} \pm 0.010$ & $\textbf{0.782} \pm 0.008$ & $\mathbf{0.744} \pm 0.006$ & $\textbf{0.761} \pm 0.004$ \\
				\bottomrule
			\end{tabular}
		}
		\caption{Comparative detection scores (AUC) in A(S)DI modes. Dataset names (stars) and parallactic amplitude $\Delta_{\phi}$ are reported on top. ASDI mode (which is better than ADI) is not supported by MODEL\&CO.}
		\label{table:main}
	\end{center}
	\vspace*{-8mm}
\end{table*}

\subsection{Quantitative and qualitative evaluations}

\paragraph{Statistical model.}
We evaluate the impact of leveraging multiple spatial scales and symmetries in our statistical model by testing its detection performance across configurations listed in Table~\ref{table:ablations}. Results show that using both symmetries and scales is crucial in ADI, where statistical parameters suffer from self-subtraction without the robustness of added spectral diversity.
\begin{figure}[t]
	\includegraphics[width=0.47\textwidth]{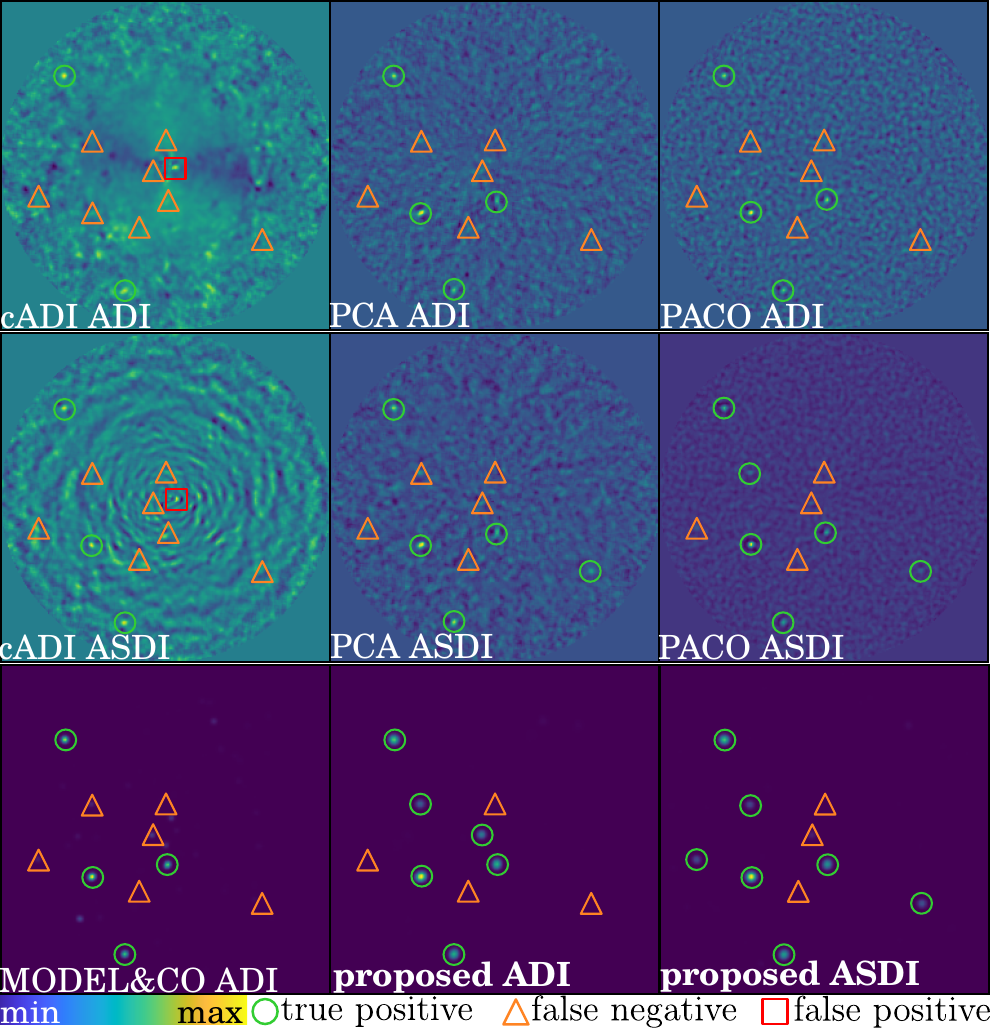}
	\caption{Detection maps on observations of HD 159911 star with synthetic exoplanets. The (calibrated) detection threshold is equivalent for all methods. The proposed approach here detect 1 additional source compared to the second best method (PACO ASDI).}
	\label{fig:samples}
\end{figure}
We then evaluate the statistical model's performance in estimating flux and sub-pixel position (regression tasks) using the optimization process from Sec. \ref{subsec:statistical_model}. Table~\ref{table:charac} shows average errors across 1,351 synthetic exoplanets detectable by both considered methods. Our approach outperforming others in almost all cases.

\vspace*{-0.5cm}
\paragraph{End-to-end learnable model.}
We evaluate detection performance on 5 observations using synthetic exoplanet injections through direct models \eqref{eq:adi_direct_model}-\eqref{eq:asdi_direct_model}. For each observation, 100 cubes are generated, totaling 1,000 injected exoplanets. This process is repeated 5 times with different random seeds, and AUC scores are reported in Table~\ref{table:main}. Examples of representative detection maps are shown in Fig.~\ref{fig:samples}. The proposed approach matches or outperforms state of the art algorithms in ADI mode. In ASDI mode, it shows a significant boost in detection sensitivity, consistently surpassing comparative methods.

\begin{figure}[t!]
	\centering
	\includegraphics[width=0.47\textwidth]{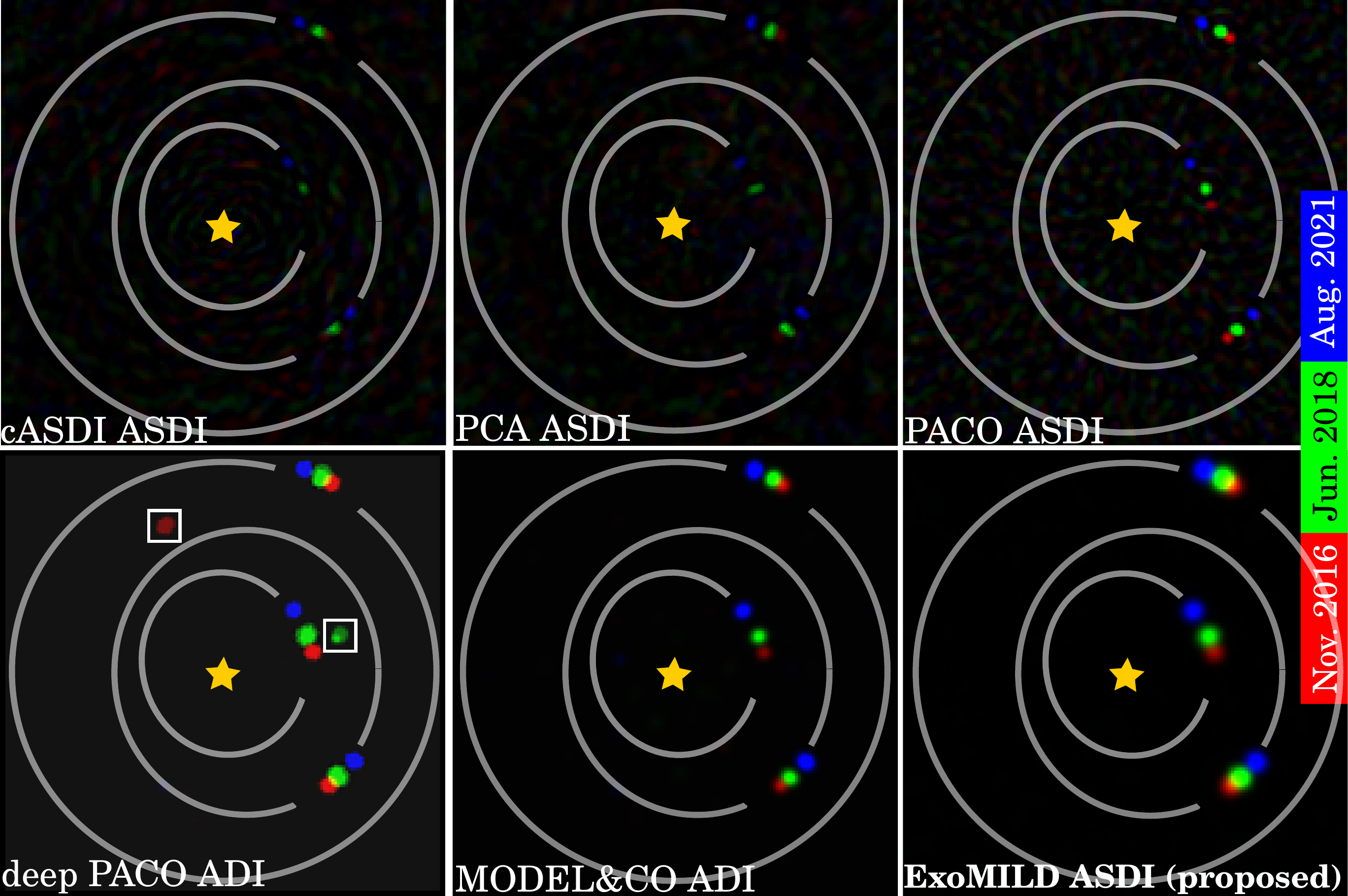}
	\caption{Detection maps on 3 observations (stacked in false RGB colors) of HR 8799 star.
		The elliptical arcs depict the estimated (projected) orbits of three known exoplanets, with the detection results shown as red, green and blue dots for the corresponding 2016, 2018 and 2021 observations.
		Squares are for false alarms identified in \cite{bodrito2024model}, Fig. 17.}
	\label{fig:real}
\end{figure}

\vspace*{-0.5cm}
\paragraph{Validation on real data.}
\label{subsec:real_data}
We compare detection methods using 3 observations, spanning several years, of the HR 8799 star. The stacked detection maps, highlighting exoplanets' orbital motion, are shown in Fig.~\ref{fig:real}.
All detection maps use a consistent unit (signal-to-noise score) and dynamic range. The proposed method maximizes detection confidence with no false alarms.

%% file: sec/conclusion.tex
\section{Conclusion}
\label{sec:conclusion}

We propose a novel hybrid approach for exoplanet imaging that combines a multi-scale statistical model with deep learning, capturing spatial correlations in the nuisance component for simultaneous detection and flux estimation. This approach provides statistically grounded detection scores, unbiased estimates, and native uncertainty quantification. Tested on VLT/SPHERE data, it outperforms state-of-the-art techniques, demonstrating efficiency and robustness across varied data qualities. 
Its versatility and reduced computational complexity make it ideal for large-scale surveys. The approach will be extended to handle higher spectral resolution data. Additionally, the nuisance model can be adapted for reconstructing spatially extended objects like circumstellar disks, namely birthplaces of exoplanets.

\section*{Acknowledgments}

\noindent This work was supported by the French government under management of Agence Nationale de la Recherche as part of the “France 2030" program, PR[AI]RIE-PSAI projet (reference ANR-23-IACL-0008) and MIAI 3IA Institute (reference ANR-19-P3IA-0003), 
and by the European Research Council (ERC) under grant agreement 101087696 (APHELEIA project). 
This work was also supported by the ERC under the European Union's Horizon 2020 research and innovation programme (COBREX; grant agreement 885593), the ANR under the France 2030 program (PEPR Origins, reference ANR-22-EXOR-0016), the French National Programs (PNP and PNPS), and the Action Spécifique Haute Résolution Angulaire (ASHRA) of CNRS/INSU co-funded by CNES.
This work was granted access to the HPC resources of IDRIS under the allocation 2022-AD011013643 made by GENCI.
JP was supported in part by the Louis Vuitton/ENS chair in artificial intelligence and a Global Distinguished Professorship at the Courant Institute of Mathematical Sciences and the Center for Data Science at New York University.

%% file: sec/supplementary.tex
\clearpage
\setcounter{page}{1}
\maketitlesupplementary
\appendix

\section{Technical details}

\subsection{Estimation of the statistical parameters}
\label{sec:estimation_params}

In this section, we provide additional information regarding the estimation of the statistical parameters of the Gaussian distribution (see Sec. \ref{subsec:statistical_model}).
These parameters are estimated by maximum likelihood:
\begin{equation} \label{eq:params_mean_S}
\widehat{\V m}_j = \frac{1}{T} \sum_t \V y_t^{(j)},\, \widehat{\M S}_j = \frac{1}{T} \sum_t \big(\V y_t^{(j)} - \widehat{\V m}_j \big) \big(\V y_t^{(j)} - \widehat{\V m}_j \big)^\top,
\nonumber
\end{equation}
where $\widehat{\M S}_j$ in $\mathbb{R}^{p \times p}$ is the sample covariance matrix.
Given that~$\widehat{\M S}_j$ usually suffers from a large variance due to small sample statistics (i.e., $p  \ge T \simeq 70$), we use a shrinkage estimator leading to the final covariance estimator $\widehat{\M C}_j$:
\begin{equation} \label{eq:params_C}
  \widehat{\M C}_j = (1 - \widehat{\rho}_j) \, \widehat{\M S}_j + \widehat{\rho}_j \, \widehat{\M D}_j \,,
\end{equation}
\noindent where $\widehat{\M D}_j$ in $\mathbb{R}^{p \times p}$ is the low-variance/high-bias diagonal matrix formed with diagonal elements of $\widehat{\M S}_j$, and $\widehat{\rho}_j$ in $[0,\, 1]$. Hyper-parameter $\widehat{\rho}_j$ can be estimated optimally in a data-driven fashion by risk minimization between the true (unknown) covariance $\M C_j$ and its shrunk counterpart $\widehat{\M C}_j$. This leads to the closed-form expression \citep{flasseur2024shrinkage}:
\begin{equation} \label{eq:shrinkage_rho}
  \widehat{\rho}_j = \frac{\tr( \widehat{\M S}^2_j ) + \tr^2(\widehat{\M S}_j) - 2 \tr( \widehat{\M S}_j \circ \widehat{\M S}_j )}{( T + 1 ) ( \tr( \widehat{\M S}_j^2) - \tr( \widehat{\M S}_j \circ \widehat{\M S}_j ) )}\,,
\end{equation}
where $\circ$ stands for Hadamard (element-wise) product.

\subsection{Dense detection map computation details}
\label{sec:dense_map}

In this section, we provide additional details on the computation of the dense detection map introduced in Sec. \ref{subsec:statistical_model}. 
To obtain a full-frame detection map, $\widehat{\alpha}$ and $\widehat{\sigma}_\alpha$ must be computed for all spatial coordinates, which can be formalized by:
\begin{equation} \label{eq:min_grid}
\forall \V x_i \in \mathcal{G}_p, \ \widehat{\alpha}_i= \argmax_{\alpha_i} \ell(\alpha_i, \V x_i) \,.
\end{equation}
Note that it corresponds to solve independent optimization problems at each location $i$.
To efficiently solve Eq.~\eqref{eq:min_grid}, we employ a fast approximation to compute the terms $a(\V x_t)$ and $b_t(\V x_t)$ at each time step and for every initial position.
First, the terms $a_t$ and $b_t$ are computed for all positions in the grid $\mathcal{G}_p$ of spatial pixels.
In the following, we note $b_{ti} = b_{t}(\V x_i)$, $a_i = a(\V x_i)$ and $\V h_{ji} = \V h^{(j)}(\V x_i)$ for all $\V x_i$ in $\mathcal{G}_p$, such that:
\begin{align}
  \forall (i, t), b_{it} &= \sum_{j \in S(\V x_i)} w_j \, \V h_{ji}^\top \, \widehat{\M C}_{j}^{-1} \, \left( \V y_t^{(j)} - \widehat{\V m}_{j} \right) \,, \label{eq:b}\\
  \forall i, a_i &= \sum_{j \in S(\V x_i)} w_j \, \V h_{ji}^\top \, \widehat{\M C}_{j}^{-1} \, \V h_{ji}\,.\label{eq:a}
\end{align}
With our convolutional approach, these terms can be computed efficiently by Cholesky factorization of the precision matrix, as detailed in Sec. \ref{sec:conv_ab}. 
Let $\V b_t$ in $\mathbb{R}^{H\times W}$ and $\V a$ in $\mathbb{R}^{H\times W}$ represent the resulting quantities.
Then, $b_t(\V x_t)$ and $a(\V x_t)$ can be approximated by spatial interpolation of $\V b_t$ and $\V a$.
However, the trajectories of all exoplanets in the observation are governed by a common parallactic rotation vector $\V \phi$ in $\mathbb{R}^T$.
So in practice, this interpolation can be computed efficiently for each time-step $t$, by rotating the frames $\V a$ and $\V b_t$ by an angle $-\phi_t$.
Thus, a dense detection map $\widehat{\V \gamma}$ in $\mathbb{R}^{H\times W}$ can be obtained with:
\begin{equation} \label{eq:detection_score_fast}
\widehat{\V \gamma} =\frac{\widehat{\V \alpha}}{\widehat{\V \sigma}} = \frac{\sum_t \M R_{-\phi_t} (\V b_t)}{\sqrt{\sum_t \M R_{-\phi_t} (\V a)}} \,,
\end{equation}
\noindent where $\M R_{- \phi_t}: \mathbb{R}^{H \times W} \to \mathbb{R}^{H \times W}$ is the frame rotation operator with angle $- \phi_t$, and $\widehat{\V \alpha}, \widehat{\V \sigma}$ in $\mathbb{R}^{H \times W}$ such that:
\begin{equation} \label{eq:minimizer}
\widehat{\V \alpha} = \frac{\sum_t \M R_{-\phi_t} (\V b_t)}{\sum_t \M R_{-\phi_t} (\V a)} ~~~\text{and}~~~
\widehat{\V \sigma} = \frac{1}{\sqrt{\sum_t \M R_{-\phi_t} (\V a)}} \,.
\end{equation}

\subsection{Fast computation of $\V b_{t}$ and $\V a$} \label{sec:conv_ab}

In this section, we detail an efficient strategy to derive a dense two-dimensional detection map $\widehat{\V \gamma}$ serving as the final quantity to detect exoplanets, see Sec. \ref{subsec:statistical_model}. 

The terms $\V b_t$ and $\V a$ in $\mathbb{R}^{H \times W}$ involved in $\widehat{\V \gamma}$ (see Eqs.~\eqref{eq:b}--\eqref{eq:detection_score_fast}) can be computed efficiently through:
\begin{align}
\V b_t &= \sum_j w_j \mathbf{Q}_j (\V b^{(j)}_t) \,, \\
\V a &= \sum_j w_j \mathbf{Q}_j (\V a^{(j)}) \,, 
\end{align}
\noindent where $\mathbf{Q}_j$ is the linear operator that places the patch indexed by $j$ in the correct position in the image, and $\V b_t^{(j)}$, $\V a^{(j)}$ are in $\mathbb{R}^p$. The individual terms $\V a^{(j)}$ and $\V b_t^{(j)}$ are given by:
\begin{align}
\V b^{(j)}_t &= \M P^\top \widehat{\M C}_{j}^{-1} \, \left( \V y_t^{(j)} - \widehat{\V m}_{j} \right) \,, \label{eq:bj}\\
\V a^{(j)} &= \text{diag}\left(\M P^\top \widehat{\M C}_{j}^{-1} \M P \right) \label{eq:aj}\,,
\end{align}
\noindent where $\M P$ in $\mathbb{R}^{p \times p}$ is a circulant matrix, with each column representing the flattened PSF centered on a different pixel.
Equation~\eqref{eq:bj} can be implemented efficiently with a convolution operation.
Implementing Eq.~\eqref{eq:aj} directly is however computationally challenging 
as it requires storing the matrix $\M P$, which is highly inefficient, especially for large patches.
To circumvent this issue, we propose to decompose $\V a^{(j)}$ as:
\begin{equation} \label{eq:aj_cholesky}
\V a^{(j)} = \mathbf{N} (\widehat{\M L}_j \M P)\,,
\end{equation}
\noindent where $\mathbf{N}: \mathbb{R}^{p \times p} \to \mathbb{R}^{p}$ is the operator that computes the column-wise squared $\ell_2$--norm, 
and the matrix $\widehat{\M L}_j$ in $\mathbb{R}^{p \times p}$ is the lower triangular matrix obtained via Cholesky decomposition of the precision matrix (i.e., $\widehat{\M C}_{j}^{-1} = \widehat{\M L}_j \widehat{\M L}_j^\top$).
The formulation of Eq.~\eqref{eq:aj_cholesky} allows implementing the matrix multiplication with $\M P$ efficiently as a convolution operation.

\subsection{Iterative optimization scheme for astrometry} \label{sec:iteration_astrometry}

We detail in this section the characterization procedure mentioned in Sec.~\ref{subsec:statistical_model} for precisely estimating the flux $\alpha$ and the two-dimensional sub-pixel position $\V x_0$ of an exoplanet.
This procedure is described by \cref{alg:charac} in pseudo-code form, where $\V z = [\alpha, \V x_0] $ in $\mathbb{R}^3$ is the variable to be refined. 
 The position is initialized to the coordinates of the pixel with the highest detection score from the detection step, while the flux is initialized using the value extracted from the flux map $\V \alpha$ at that location. 
 We decompose the involved loss function $\ell$ as:
 \begin{equation}
 \ell(\V z, \{\V m_j, \M C_j \}_j) = \sum_t \sum_{j \in S(\V x_t)} w_j \ell_j(\V z, \V m_j, \M C_j)\,,
 \end{equation}
\noindent where:
\begin{multline}
\ell_j([\alpha, \V x_0], \V m_j, \M C_j) = \\ \hfill \norm{ \M L_j \left(\V y_t^{(j)} - \alpha \V h^{(j)}(r(\V x_0, \phi_t)) - \V m_j \right) }_2^2 \,,
\end{multline}
such that $\M C_j^{-1} = \M L_j \M L_j^\top$.
We also define the feasible set $Z$ such that $\alpha > 0$ (enforcing a non-negativity constraint on the exoplanet flux) and the estimated position remains within a fixed radius of the initial estimate. We denote by $\text{Proj}_Z$ the projection associated with it.
In \cref{alg:charac},  we also denote by $\text{Estim(.)}$ the procedure described in Eqs.~\eqref{eq:params_mean_S}--\eqref{eq:shrinkage_rho} to estimate the statistical parameters of the nuisance component.

\begin{algorithm}[t!]
\caption{Iterative procedure for flux and position estimation.}\label{alg:charac}
\begin{algorithmic}
\Require $\V y \in \mathbb{R}^{T \times H \times W}$, $\V z \in \mathbb{R}^3$
\For{$l = 0 \cdots L-1$}
    \State $\alpha, \V x_0  \gets \V z$
    \For{$t = 0 \cdots T-1$}
        \Comment{Update nuisance estimate}
        \State $\V x_t \gets r(\V x_0, \phi_t)$
        \State $\V s_t \gets \V y_t - \V h(\V x_t) $
    \EndFor
    \State $\V g, \M H \gets \M 0_{\mathbb{R}^3}, \M 0_{\mathbb{R}^{3 \times 3}}$
    \For{$t = 0 \cdots T-1$}
        \Comment{Gradient and Hessian}
        \For{$j \in S(\V x_t)$}
            \State $\widehat{\V m}_j, \widehat{\M C}_j \gets \text{Estim}( \{\V s^{(j)}_t\}_{t})$
            \Comment{Eqs.~\eqref{eq:params_mean_S}--\eqref{eq:shrinkage_rho}}
            \State $\V g \gets \V g + w_j \frac{\partial \ell_j}{\partial \V z}(\V z, \widehat{\V m}_j, \widehat{\M C}_j)$
            \State $\M H \gets \M H + w_j \frac{\partial^2 \ell_j}{\partial \V z^2}(\V z, \widehat{\V m}_j, \widehat{\M C}_j)$
        \EndFor
    \EndFor
    \State $\V z \gets \text{Proj}_Z( \V z - \M H^{-1} \V g$) \Comment{Projection on feasible set}
\EndFor
\State \Return{ $\V z$}
\end{algorithmic}
\end{algorithm}

\subsection{Prior information on the exoplanet spectrum}

The channel-wise detection maps $\widetilde{\V \gamma}_c = \widetilde{\V b}_c / \sqrt{\widetilde{\V a}_c}$ are combined into a single detection map by averaging $\widetilde{\V a}_c$ and $\widetilde{\V b}_c$ with weights $e_c$ in $\mathbb{R}^+$ satisfying $\sum_c e_c = 1$ such that:
\begin{equation}
\widetilde{\V a}_C = \sum_c e_c \, \widetilde{\V a}_c \, \quad \text{and} \quad
\widetilde{\V b}_C = \sum_c e_c \, \widetilde{\V b}_c \,.
\end{equation}
The final detection score, spectrally combined over $C$ channels, is $\widetilde{\V \gamma}_C = \widetilde{\V b}_C / \sqrt{\widetilde{\V a}_C}$ in $\mathbb{R}^{H \times W}$.
The weights $e_c$ reflect prior assumptions about the exoplanet spectrum, and can be used (if available) to boost the detection of exoplanets having similar spectrum, see e.g., \cite{flasseur2020paco,chomez2023preparation}. 
For all experiments and baselines, we assumed a non-informative flat prior (i.e., $e_c = 1 / C$). If additional information about the exoplanet’s spectrum is available, the prior can be adjusted accordingly.

\subsection{Calibration procedure}
\label{sec:calibration}

\begin{figure}[t!]
\centering
\includegraphics[width=0.4\textwidth]{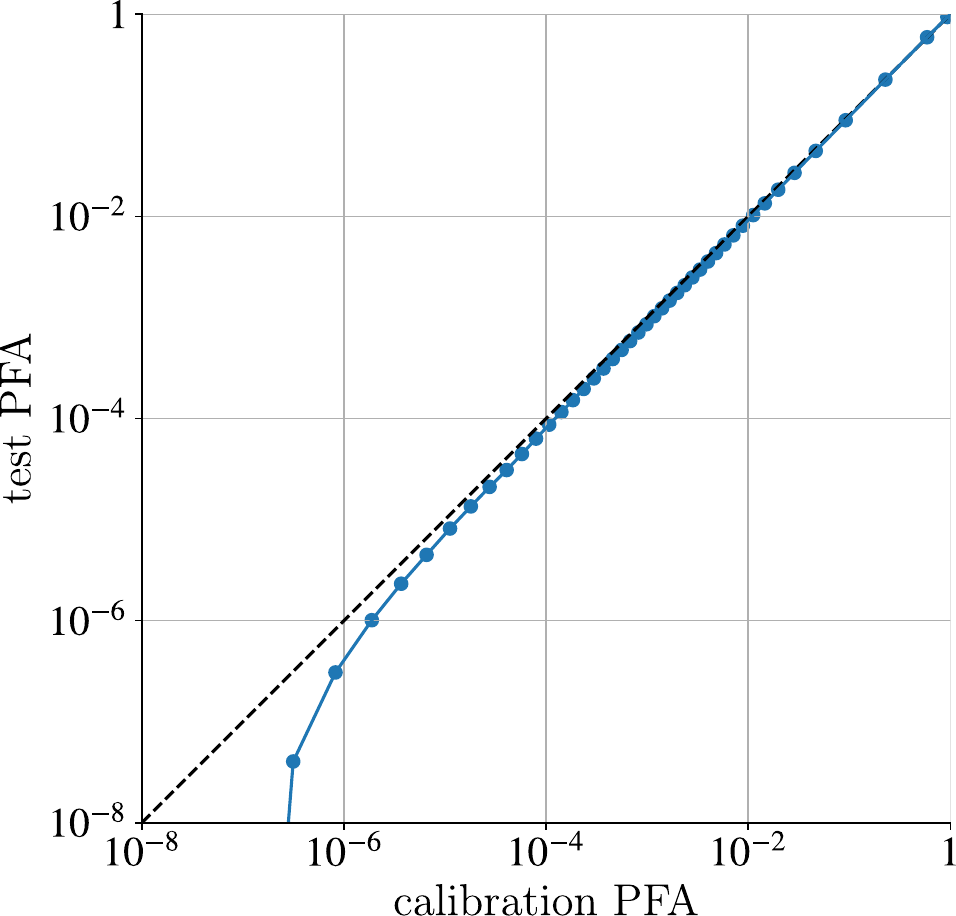}
\caption{Calibrated PFA  versus test PFA: the calibration procedure allows obtaining a reliable approximation of the PFA, except in the extreme low-PFA regime (high detection score under $\mathcal{H}_0$) where samples are exceedingly rare.}
\label{fig:calib}
\end{figure}

This section outlines the calibration procedure for the full model described in \cref{subsec:learning}, following the approach proposed in \cite{bodrito2024model}, which we briefly summarize here.

\begin{figure*}[t!]
     \centering
     \begin{subfigure}[b]{0.3\textwidth}
         \centering
         \includegraphics[width=\textwidth]{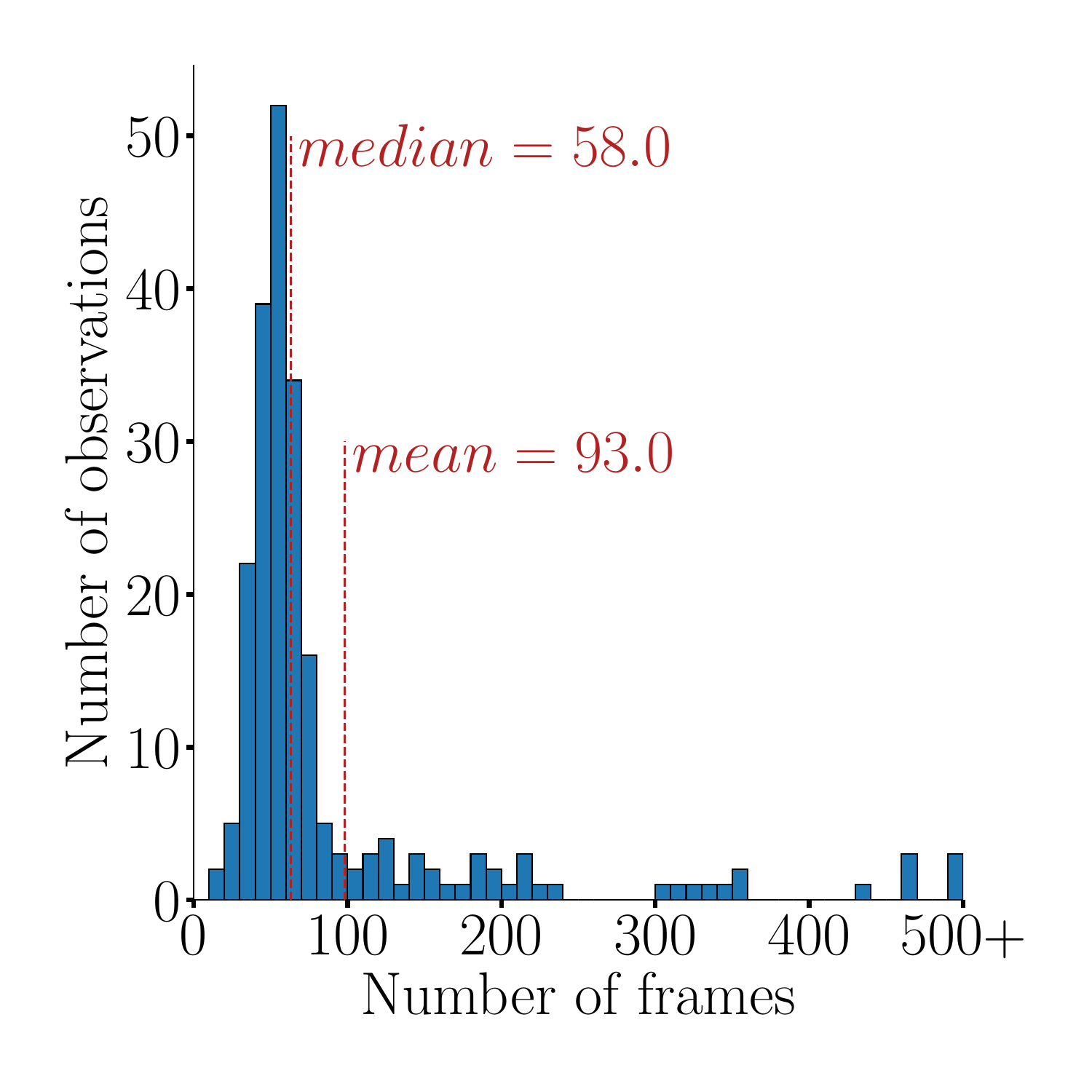}
         \caption{Distribution of the number of frames per observation}
     \end{subfigure}
     \hfill
     \begin{subfigure}[b]{0.3\textwidth}
         \centering
         \includegraphics[width=\textwidth]{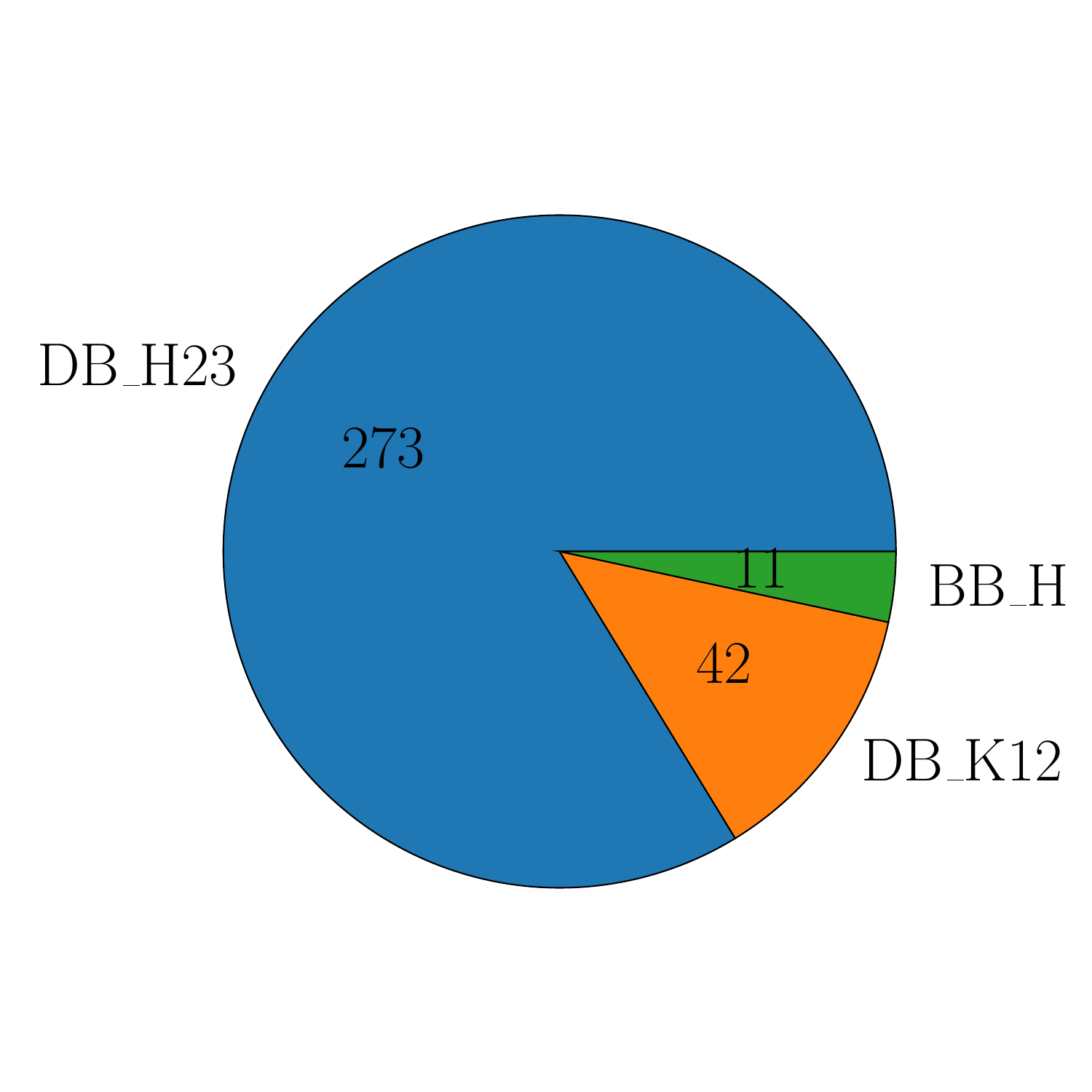}
         \caption{Repartition of the filters used}
     \end{subfigure}
     \hfill
     \begin{subfigure}[b]{0.3\textwidth}
         \centering
         \includegraphics[width=\textwidth]{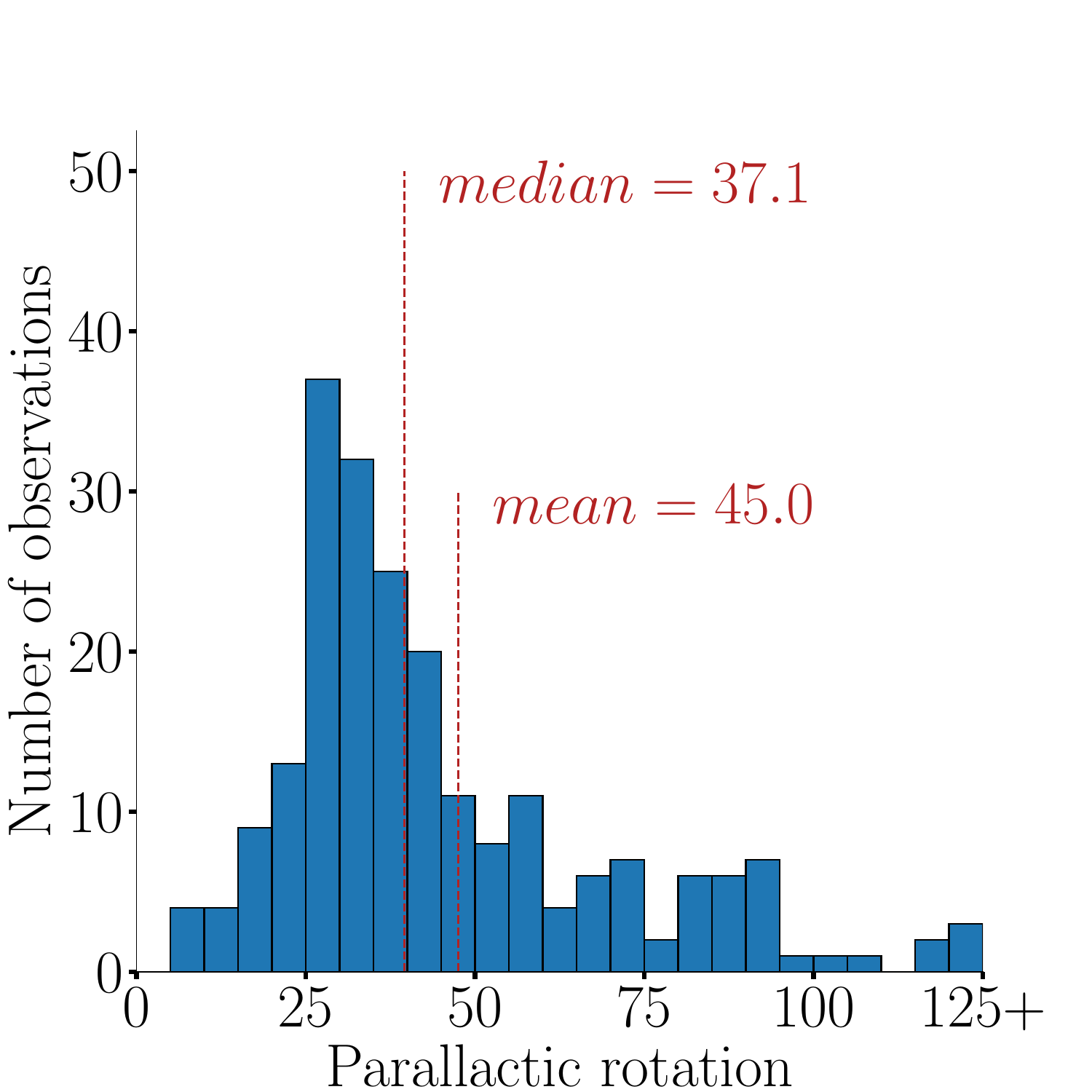}
         \caption{Distribution of the parallactic rotation per observation}
     \end{subfigure}
        \caption{Additional details on the F150 dataset}
        \label{fig:dataset_details}      
\end{figure*}

We aim to calibrate the predicted probability of false alarm (PFA), a critical metric for astronomers to quantify detection uncertainty. 
This requires characterizing the output distribution of our method under the null hypothesis $\mathcal{H}_0$ (i.e., when no source is present), which we denote $\widetilde{\gamma}|\mathcal{H}_0$. For that purpose, we use a calibration dataset $\mathcal{D}_\text{calib}$, consisting of 15 observations excluded from training, validation, and testing. 
Temporal shuffling of the frames and parallactic rotation reversal are applied to augment the dataset and ensure no real source is present. These custom data-augmentations disrupt the temporal coherence of any (potentially unknown) real exoplanets, preventing them from biasing the training and calibration processes. 
We apply the model to 10,000 cubes from $\mathcal{D}_\text{calib}$, concatenate all output pixels into a single vector $\widetilde{\V \gamma}_\text{calib}$ of length $n_\text{calib}$, and compute the empirical cumulative distribution function (eCDF):
\begin{equation}
  \forall \, \tau \in \mathbb{R}, \quad \widehat{F}_{\widetilde{ \V \gamma } | \mathcal{H}_0}(\tau) := \frac{1}{n_{\text{calib}}} 
  \sum_{i=0}^{n_{\text{calib}} -1} {\M 1}_{\widetilde{\V \gamma}_{\text{calib}, i} \leq \tau} \,.
\end{equation}
\noindent where $\M 1$ is the indicator function. 
The estimated probability of false alarm is related to the eCDF through:
\begin{equation}
  \forall \, \tau \in \mathbb{R}, \quad \widehat{\text{PFA}}(\tau) = \widehat{\mathbb{P}}(\widetilde{ \V \gamma  } > \tau | \mathcal{H}_0) 
  = 1 - \widehat{F}_{\widetilde{ \V \gamma } | \mathcal{H}_0}(\tau)\,.
\end{equation}
In \cref{fig:calib}, we evaluate the PFA across a range of detection thresholds using the calibration set and apply the same thresholds to the test set for comparison. The results show that the calibrated PFA closely matches the test PFA, except in the extremely low-PFA regime. This slight discrepancy is expected, as high detection scores under $\mathcal{H}_0$ are exceedingly rare, leading to increased noise in PFA estimation. In this regime, our estimate is nevertheless conservative, i.e., the experienced PFA is slightly smaller than predicted one.

\subsection{Implementation details}

Our model implementation and training are based on the PyTorch framework. 
We optimize the neural network weights using the Adam optimizer with a learning rate of $5\times 10^{-4}$, a batch size of 16, and 50,000 iterations. 
We apply data augmentation techniques, including random $90\degree$ rotations, vertical/horizontal flipping, and frame shuffling. 
Bicubic interpolation is used for spectral and exoplanet alignment. 
For ADI, the model is trained on 64 frames, while for ASDI, data cubes have 32 frames with two channels each.
All evaluations and tests are done with 64 frames in both ADI and ASDI.
To ensure accurate results, pixels containing real known exoplanets are masked out during training before the temporal summation in \cref{eq:minimizer}. The loss is calculated on the final output for pixels with at least 8 valid frames, unaffected by real exoplanets.

\section{Dataset description}

SHINE-F150 is a subset of the SPHERE large survey, with 322 data cubes ($20,462$ images in total) from 150 stars. 
The spatial resolution is 12 milliarcseconds/pixel. The spectral resolution $\lambda / \Delta_\lambda$ is 20. Detailed description is
in [39].
We followed [5] to exclude corrupted data, leaving 220 datasets, with 165 used for training. 
The main statistics of the F150 dataset are shown in Fig.~\ref{fig:dataset_details}.

\section{Comparison of state-of-the-art methods}

\begin{table}[H]
	\begin{center}
  \resizebox{\columnwidth}{!}{%
		\begin{tabular}{c|cccccccc}
			\toprule
			\textbf{Method} & \textbf{nuisance model} & \textbf{P1} & \textbf{P2} & \textbf{P3} & \textbf{P4} & \textbf{P5} & \textbf{P6} & \textbf{P7}\\
			\midrule
			PACO & statistical on data patches & \redcross & \greencheck & \greencheck & \greencheck & \redcross &  \greencheck &  \greencheck\\
			SODINN & learned on PCA residuals & \redcross & \redcross & \redcross & \greencheck \greencheck & \redcross & \redcross & \redcross \\
			deep PACO & learned after PACO whitening & \redcross & \redcross & \greencheck & \greencheck \greencheck & \redcross & \greencheck & \redcross\\
			Super-RDI & PCA & \greencheck & \greencheck & \redcross & \greencheck \greencheck & \redcross & \redcross & \greencheck \\
			MWIN5-RB & learned on PCA residuals & \greencheck & \greencheck & \redcross & \greencheck \greencheck & \greencheck & \redcross & \greencheck \greencheck\\
			ConStruct & generative, learned on data & \greencheck & \greencheck & \redcross & \greencheck \greencheck & \greencheck & \redcross & \greencheck \greencheck \greencheck\\			
			MODEL\&CO & statistical on learned features & \greencheck & \redcross & \redcross & \greencheck \greencheck \greencheck & \greencheck \greencheck & \redcross & \greencheck \greencheck \greencheck \\
			\textbf{proposed} & multi-scale stat. on learned features & \greencheck & \greencheck & \greencheck & \greencheck \greencheck \greencheck & \greencheck \greencheck \greencheck & \greencheck & \greencheck \greencheck \greencheck\\		
			\bottomrule
		\end{tabular}
  }
            \captionsetup{skip=0pt} 
		\caption{
        Comparison of advanced exoplanet imaging algorithms based on key desirable properties.
        P1=observation-independence of the model, P2=physical interpretability, P3=multi-spectral, P4=detection sensitivity, P5=accuracy near the star, P6=position and flux estimation, P7=practicality on large-scale surveys.}
        \label{tab:algos_properties}
	\end{center}
\end{table}

Table~\ref{tab:algos_properties} compares the key properties of state-of-the-art algorithms for exoplanet detection via direct imaging.

\section{Additional results}

In this section we provide additional visual results, and show detection maps obtained with synthetic exoplanets.
Consistent with the experimental findings in the main body of the paper, our method performs on par or better than baselines in ADI, and outperforms other approaches in ASDI.

\clearpage

\vspace*{0.001mm}
\begin{figure}[h]
\centering
\includegraphics[width=0.47\textwidth]{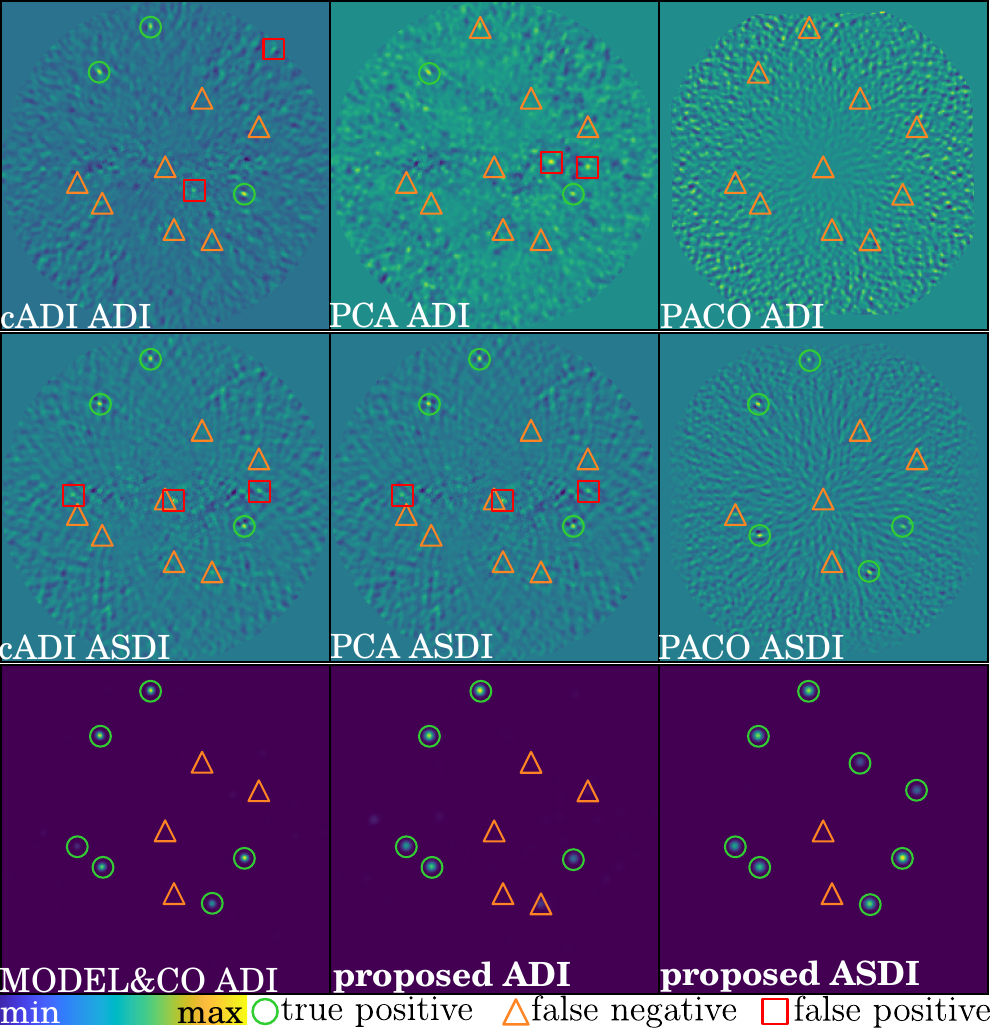}
\caption{Detection maps on observations of HD 102647 star with synthetic exoplanets.}
\label{fig:samples}
\end{figure}

\vspace*{8.75mm}
\begin{figure}[h]
\centering
\includegraphics[width=0.47\textwidth]{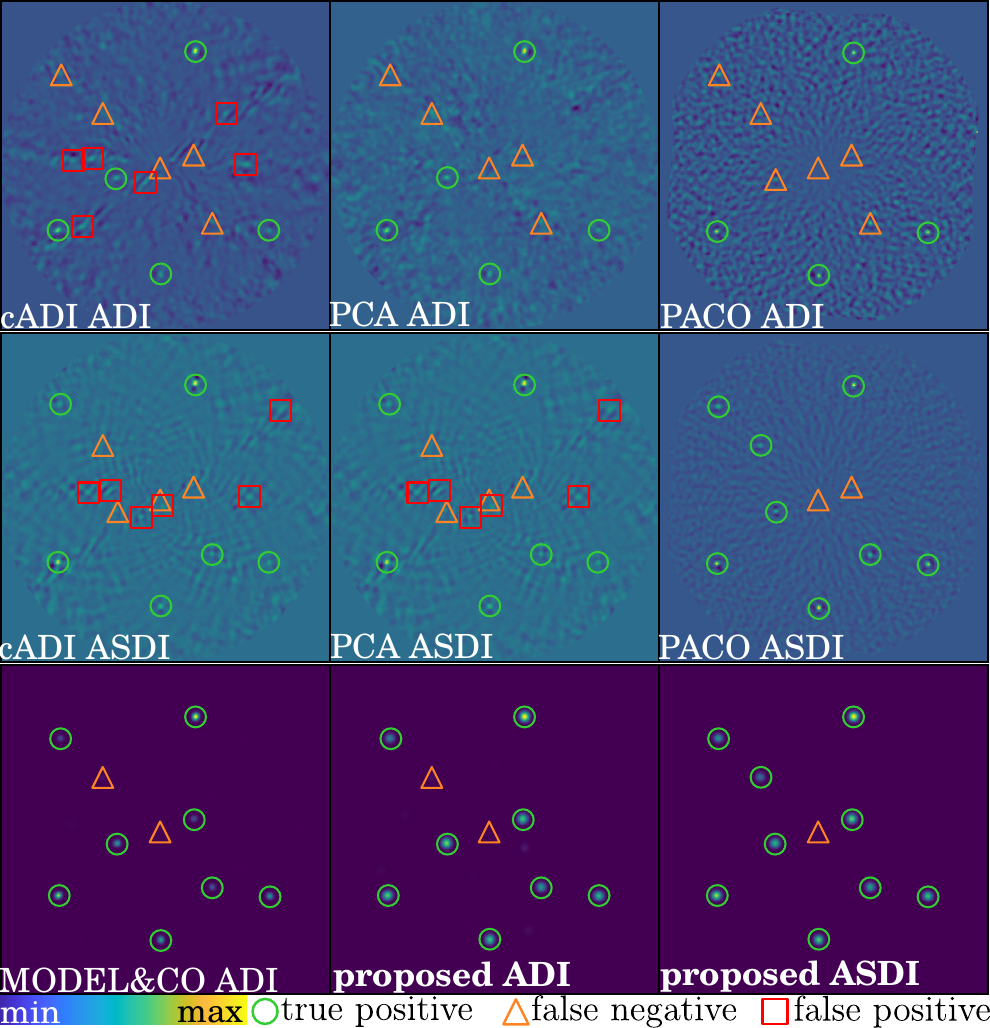}
\caption{Detection maps on observations of HD 188228 star with synthetic exoplanets.}
\label{fig:samples}
\end{figure}

\begin{figure}[h]
\centering
\includegraphics[width=0.47\textwidth]{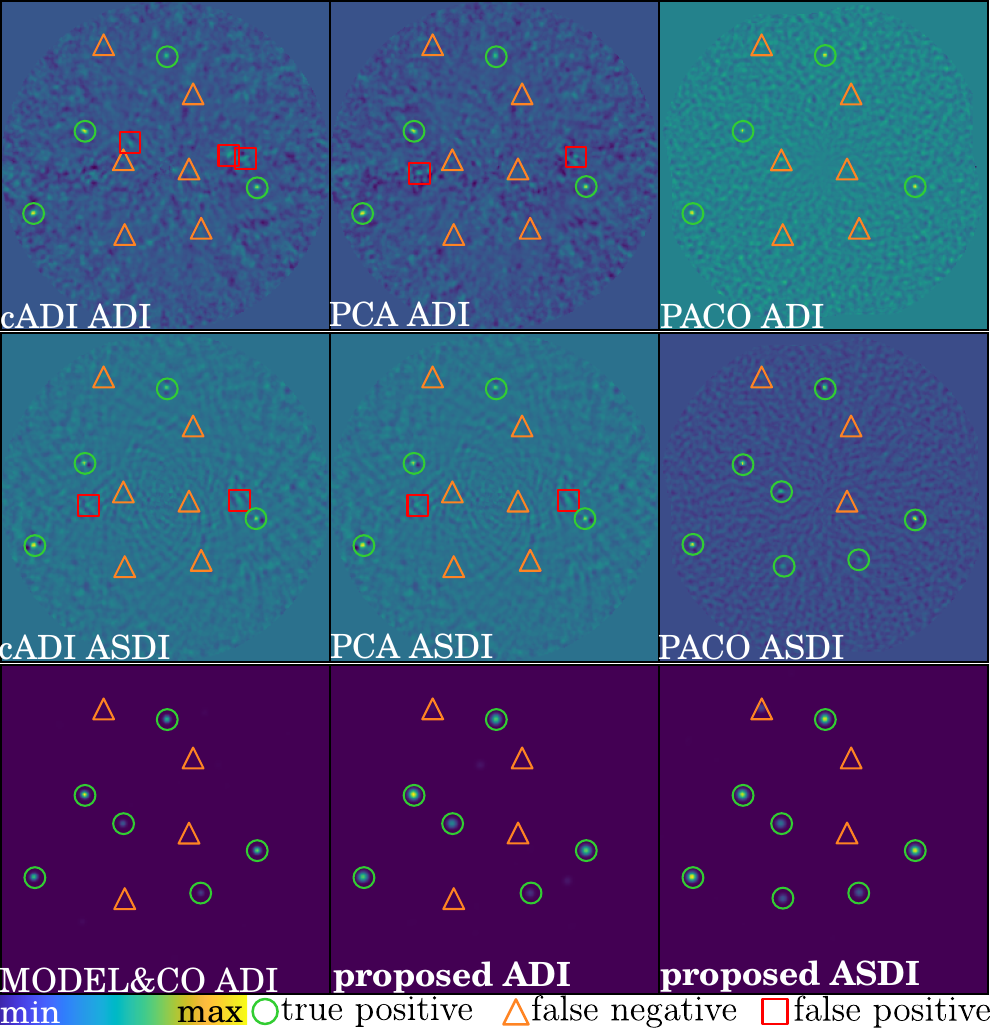}
\caption{Detection maps on observations of HD 206860 star with synthetic exoplanets.}
\label{fig:samples}
\end{figure}

\begin{figure}[h]
\centering
\includegraphics[width=0.47\textwidth]{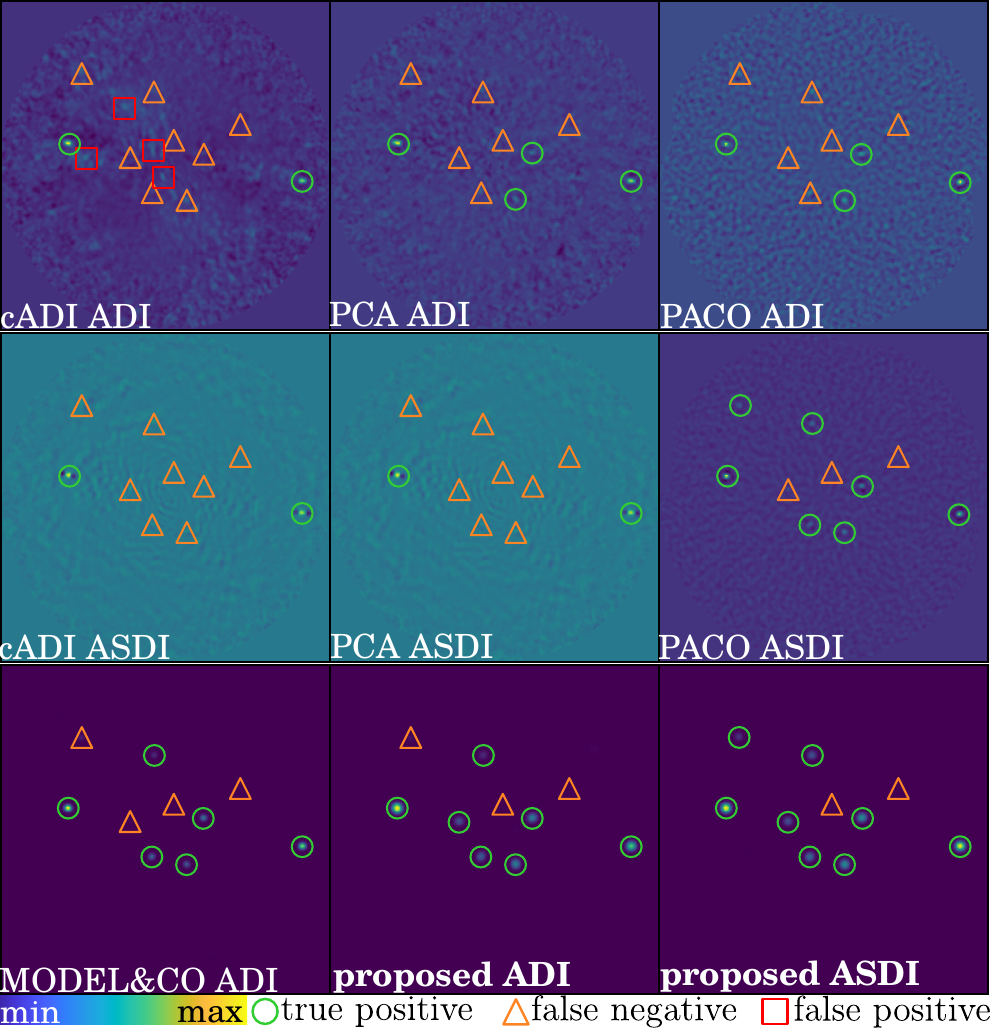}
\caption{Detection maps on observations of HD 216803 star with synthetic exoplanets.}
\label{fig:samples}
\end{figure}

\clearpage